# Graphene Heterostructure-Based Non-Volatile Memory Devices with Top Floating Gate Programming

*Gabriel L. Rodrigues[1,2], Ana B. Yoshida[1,2], Guilherme S. Selmi[1,2], Nickolas T.K.B de Jesus[1], Igor Ricardo[3], Kenji Watanabe[4], Takashi Taniguchi[5], Rafael F. de Oliveira[1], Victor Lopez-Richard[3], Alisson R. Cadore[1,6],\**

[1]Brazilian Nanotechnology National Laboratory (LNNano), Brazilian Center for Research in Energy and Materials (CNPEM), 13083-200, Campinas, Sao Paulo, Brazil
[2]"Gleb Wataghin" Institute of Physics, State University of Campinas, 13083-970, Campinas, São Paulo, Brazil
[3]Departamento de Física, Universidade Federal de São Carlos, 13565-905, São Carlos, São Paulo, Brazil
[4]Research Center for Electronic and Optical Materials, National Institute for Materials Science, 1-1 Namiki, 305-0044, Tsukuba, Japan
[5]Research Center for Materials Nanoarchitectonics, National Institute for Materials Science, 1-1 Namiki, 305-0044, Tsukuba, Japan
[6]Programa de Pós-Graduação em Física, Instituto de Física, Universidade Federal de Mato Grosso, 78060-900, Cuiabá, Mato Grosso, Brazil
**\*alisson.cadore@lnnano.cnpem.br**

**ABSTRACT**

We present a graphene-based memory platform built on dual-gated field-effect transistors (GFETs). By integrating a lithographically defined metal patch directly atop the hexagonal boron nitride (hBN)-graphene channel, the device functions simultaneously as a top gate, floating gate (FG) reservoir, and active reset contact. This architecture forms an ultrathin van der Waals heterostructure with strong capacitive coupling to the back-gate, confirmed by a dynamic model, enabling a tunable and wide memory window that scales with back-gate voltage and is further enhanced by reducing hBN thickness or increasing FG area. Our devices demonstrate reversible, high-efficiency (>90%) charge programming, robust non-volatile behavior across 10 – 300 K and a wide range of operation speeds, and endurance beyond 9800 cycles. Importantly, a grounded top electrode provides on-demand charge erasure, offering functionality that is absent in standard FG designs. These results position hBN/graphene-based GFETs as a compact, energy-efficient platform for next-generation 2D flash memory, with implications for multilevel memory schemes and cryogenic electronics.

**INTRODUCTION**

Flash memory has become the cornerstone of modern non-volatile storage technologies, owing to its high density, scalability, and relatively low fabrication cost.[1–6] The standard flash memory cell relies on a floating-gate (FG) architecture, where a conductive FG layer (i.e., electrically isolated by surrounding dielectrics) stores charges injected from the (semi)conductor channel through a thin tunneling dielectric.[1–6] The resulting shift in threshold voltage defines the device's logic states.[3,7–15] While reducing the thickness of the tunneling dielectric can lower program/erase voltages and improve integration density, it simultaneously increases leakage currents, undermining data retention and device reliability.[13,16–18] These trade-offs, along with fundamental material limitations, are pushing conventional silicon-based flash memory toward its scaling limits.

In response, emerging materials and alternative memory architecture are under intensive investigation.[1,3–6,11,13,19–30] In this scenario, graphene-based field-effect transistors (GFETs) encapsulated with hexagonal boron nitride (hBN) have attracted considerable attention as a next-generation platform for ultrafast and energy-efficient memory devices.[3–6,8,18,31–33] Single-layer graphene (SLG) offers exceptional carrier mobility, chemical inertness, and mechanical stability,[16,17,34] while hBN provides an atomically smooth, insulating barrier with a wide bandgap,[35–38] enabling superior interface quality and dielectric performance.[36,39–42] Together, these layered materials form ultrathin van der Waal heterostructures that support highly tunable electrostatic environments,[35,43–45] ideal for novel memory functionalities. Notably, dual-gated GFET architectures allow precise control over carrier density and electrostatic doping in the SLG channel,[40,45–47] enabling programmable memory behavior through charge trapping and electrostatic hysteresis. These features offer an attractive route to emulate FG mechanisms in ultrathin platforms. However, a systematic understanding of how key structural parameters like hBN thickness, FG geometry, and gate coupling configuration influence device performance remain incomplete.

Here we introduce hBN/SLG-based GFETs in which a single lithographically defined metal patch atop the SLG channel simultaneously acts as a conventional top-gate, as an FG charge reservoir, or as an active reset contact (**Figure 1a-c**). This dual-gated geometry creates a strong capacitive coupling between the silicon back-gate and the suspended FG, enabling a wide and tunable memory window that scales linearly with the back-gate sweep amplitude and is further amplified by reducing the top-hBN thickness or enlarging the FG area. Systematic thickness- and area-dependent studies reveal programming efficiencies above 90 %, values that surpass previous 2D-based flash demonstrations while remaining fully reversible and immune to substrate-induced hysteresis. Beyond static characteristics, the device exhibits truly non-volatile operation, in agreement to our dynamic model: the memory window is invariant from 10 K to 300 K; across four orders of magnitude in sweep rates; retains its states for over a year under ambient conditions; and endures more than 9800 program/erase cycles. Crucially, grounding the top electrode erases the stored charge on demand, providing an integrated "reset" knob absent in conventional FG schemes. Together, these results establish a scalable, compact platform for ultrafast, low-voltage 2D flash memories and offer clear design rules for future multilevel storage and cryogenic electronics.

**RESULTS AND DISCUSSION**

**Figure 1a-c** shows the schematic structure of the fabricated dual-gated memory device with its electrical connections. Detailed device preparation methods and corresponding optical

images of a representative device are provided in the Supplementary Material (SM) - Section 1. **Figure 1a** illustrates the initial device, where a fully hBN/SLG/hBN encapsulated GFET is solely back-gated ($V_{BG}$). In contrast, **Figure 1b-c** shows the same GFET after the nanofabrication of an additional electrode over the hBN/SLG/hBN region (i.e., the final device), which can operate as a top-gate bias ($V_{TG}$), a floating gate, or a reset electrode. The source-drain current ($I_{DS}$) in such GFETs can be regulated by the $V_{BG}$, $V_{TG}$, and FG, as will be shown.

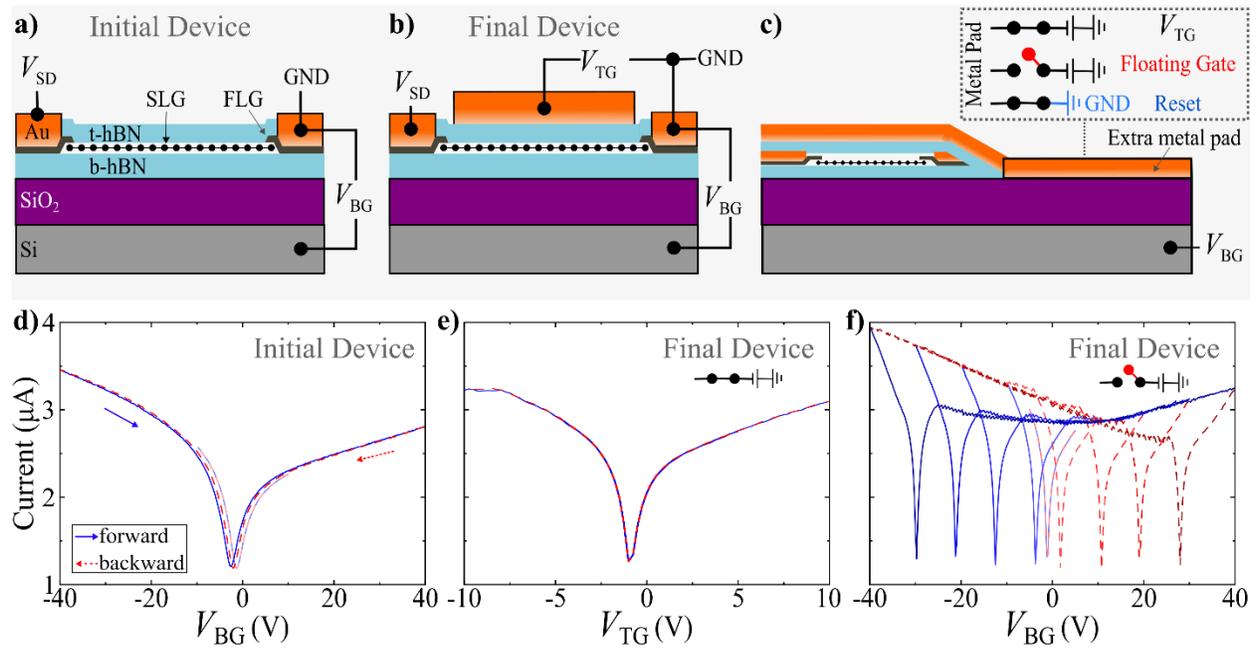

**Figure 1:** Structure of the graphene memory device. **a)** Experimental scheme of an encapsulated hBN/SLG/hBN GFET operated solely in a back-gate configuration. Cross-section view of the experimental scheme for a double-gated GFET, highlighting the **(b)** top-gate electrode and **(c)** the extra metal pad connections. The inset in (c) brings the three working principle possibilities of the top electrode: top-gate bias ($V_{TG}$); floating gate (FG); or reset. Transfer characteristics of the **(d)** initial device operated in a back-gate configuration; **e)** final device operated in a top-gate configuration, and **f)** the final device controlled by the back-gate, while the top electrode is floating. In (d), we present the transfer curves for loops of $|V_{BG, max}| = 10$ V (lighter curves) and $|V_{BG, max}| = 40$ V (darker curves), whereas in (f), $|V_{BG}|$ increases from $|V_{BG, max}| = 5$ V (lighter curves) to 40 V (darker curves). The extra metal pad size is 200 × 200 μm² in (f). Test conditions are fixed $V_{DS} = 10$ mV, same bias sweep rate, and in vacuum at 300 K.

**Figure 1d** plots the transfer curves of a typical GFET under different $V_{BG}$ loops ($|V_{BG, max}|$ = 10 V and 40V). The transfer curves of the initial device do not exhibit significant electrical hysteresis between the forward (blue lines) and backward (red lines) $V_{BG}$ sweeps. This absence of electrical hysteresis is also observed in **Figure 1e** when the same device is measured in the top-gate configuration (final device). However, a noticeable electrical hysteresis appears when the final device is measured in the back-gate configuration (**Figure 1f**), while the top-gate electrode is disconnected (floating). The hysteresis loop is observed in both $V_{BG}$ sweep directions and is modulated by the maximum $|V_{BG}|$ amplitudes, a crucial feature for memory devices. As shown in **Figure 1f**, the shift of the charge neutrality point (CNP) leads to the formation of a memory window, which can be widened as $|V_{BG}|$ increases from $|V_{BG, max}|$ = 5 V (lighter curves) to 40 V (darker curves). The shift of the SLG's CNP upon $V_{BG}$ application toward negative and positive directions corresponds to hole and electron trapping, respectively. We also measured the device performance with the FG grounded to exclude the possibility that the memory window is created by defects in the hBN flakes or in the interface of the heterostructure. SM-Section 2 provides characteristic curves in this condition, where no memory window is observed. SM-Section 3 demonstrates that the memory effect is independent of the $V_{BG}$ sweep direction, and it is reversible. SM-Section 4 brings the results for GFETs fabricated without bottom-hBN, together with the electrical characterization of devices with and without FGs. SM-Section 4 also demonstrates that the electrical hysteresis upon $V_{BG}$ loops is observed only in GFETs covered by the FG. Moreover, as will be discussed later, the retention time of the device also changes when the top electrode is grounded after the programming process, working as a reset electrode. Additionally, the memory window (i.e., described by the maximum voltage shift during dual $V_{BG}$ sweeps) is strongly influenced by the dimensions of the extra metal pad of the FG and the thickness of the top hBN layer. Thus, we can confidently infer that charges trapped in the FG electrode are responsible for the observed hysteretic behavior.

Next, we focus on the hBN/SLG/hBN heterostructure and characterize the device performance as a function of the thickness of the hBN flakes. **Figure 2a-b** shows the position of the CNP for devices with different top- and bottom-hBN thicknesses in the forward (blue symbols) and backward (red symbols) sweeps under back- and top-gate bias, respectively. From the analysis of these figures, no correlation between the memory window and hBN thickness is observed. On the contrary, **Figure 2a** shows a rather constant behavior across all thicknesses employed, regardless of the maximum $V_{BG}$ applied, while **Figure 2b** demonstrates the absence of electrical hysteresis in the $V_{TG}$ measurements. Such behavior could be associated with charge trapping at the $SiO_2$ layer, induced by electron beam (e-beam) irradiation during the lithography (EBL) of the top electrode.[48] To investigate this, we fabricated additional devices using different e-beam acceleration voltages during the EBL step. SM-Section 5 provides data for devices fabricated with 5 kV and 30 kV. The results indicate no correlation between the memory window and e-beam acceleration, showing a similar memory window for all cases. Moreover, considering that all measurements were conducted at 300 K in vacuum, we can also rule out temperature effects.[49]

The results in **Figure 1** and SM-Sections 2 and 4 prove that the memory window depends on the presence of the top electrode over the hBN/SLG/hBN stacking. Thus, it is reasonable to assume that the memory window would also be affected by the thickness of the top dielectric. **Figure 2c** plots the CNP splitting as a function of the top-hBN thickness. From this figure, one can observe an enhancement of the memory window for thinner top-hBN flakes, while the response saturates for thicker ones. It is important to mention that, so far, this saturation effect is not fully understood. The memory window, described by the ΔCNP during the dual sweep of $V_{BG}$, is extracted for representative devices and is displayed in **Figure 2d**. ΔCNP exhibits a linear relationship with the maximum $V_{BG}$ applied. The ratio between the memory window and the sweeping range, programming efficiency, can exceed 90% for a 6 nm top-hBN thickness. Notably, ΔCNP reaches approximately 72 V under $|V_{BG, max}|$ = 40 V loop, which is significantly larger than that observed in previous memory devices.[7,8,11,18] Table S1 brings a comparison of our memory devices with the literature, demonstrating that our devices show outstanding performance.

Another aspect evident from **Figure 2d** is the threshold bias required to activate the memory window. While a 6 nm thick top-hBN already exhibits electrical hysteresis at $|V_{BG}|$ = 5 V, a 45 nm thick top-hBN only displays hysteresis loops for $|V_{BG}|$ = 20 V. From the linear fit of the experimental data, it can be observed that the threshold $|V_{BG}|$ required to induce electrical hysteresis is approximately 3.3 V (6 nm top-hBN), 7.2 V (16 nm top-hBN), and 15.6 V (45 nm top-hBN), confirming the strong dependence of device operation on the top-hBN thickness. SM-Section 4 shows a similar analysis for capped hBN/SLG devices. Based on these results, achieving a non-volatile graphene memory device operating within an optimal electronic window (±5 V) would require top-hBN thicknesses smaller than 10 nm onto a 300 nm $SiO_2$ substrate. Such thicknesses have also been described as the best option for high-speed programming in 2D-based memory devices.[8,18] Therefore, the choice of an appropriate tunnelling layer thickness is a prerequisite for achieving 2D-based heterostructure flash memories.

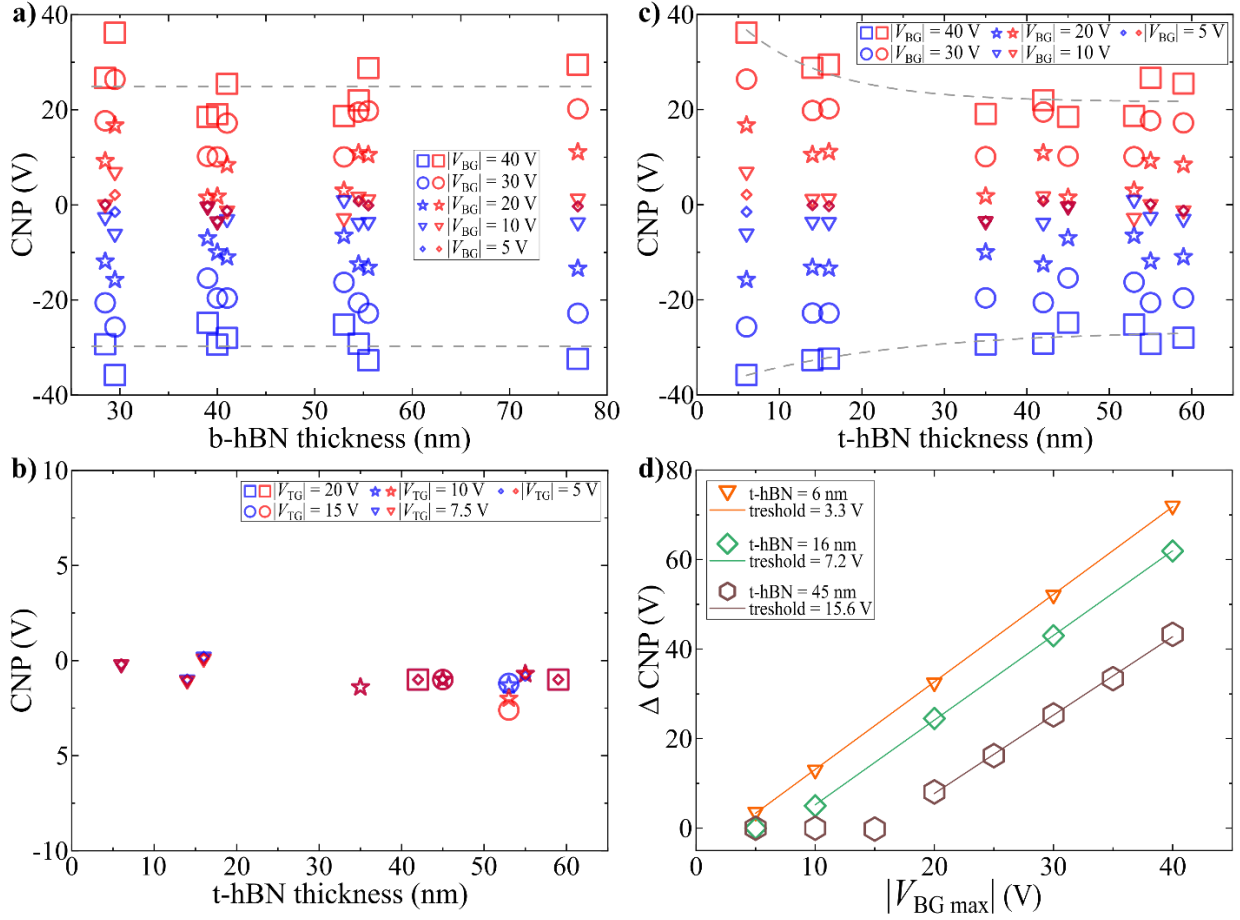

**Figure 2: a-b)** Position of the charge neutrality point (CNP) for encapsulated GFETs in the forward (blue symbols) and backward (red symbols) $V_{BG}$ and $V_{TG}$ sweeps as a function of the bottom-hBN (b-hBN) and top-hBN (t-hBN) thicknesses, respectively. Gray dashed lines illustrate no significant changes in the CNP position for different b-hBN thicknesses. **c)** CNP shift obtained during the $V_{BG}$ sweep as a function of the t-hBN thickness. Gray dashed lines serve as a guide to the eye for $|V_{BG, max}| = 40$ V data. **d)** Memory window extracted from (c) for representative devices and linear fits. Test conditions are fixed with an extra metal pad size of $200 \times 200$ μm$^2$, and in vacuum at 300 K for all GFETs.

Overall, the essence of an FG memory device is to tune the number of carriers in the floating layer by programming/erasing operations with different voltage loops or pulses.[1,5,6] When the FG electrode is located between the 2D conducting layer and the control gate[8,18], the density of the stored charge in the FG electrode after programming operations with different $|V_{BG}|$ loops has been estimated from the expression: $n = (\Delta V_{th} \times C_{blocking})/q$, where $\Delta V_{th}$, $C_{blocking}$ and $q$ are the threshold voltage shift, capacitance of the blocking dielectric, and electron charge, respectively. However, to infer the number of charges trapped in the FG of our device configuration is challenging. Wang *et al.* propose that devices like ours have three capacitances:[11] the top gate capacitance ($C_{TG}$), the back-gate capacitance ($C_{BG}$), and the coupling capacitance between the back and the metal extra pad ($C_{BG\text{-}Pad}$); giving an effective total back-gate capacitance $C_{Total} = C_{BG} + (C_{TG}^{-1} + C_{BG-Pad}^{-1})^{-1}$. More recently, we have expanded this idea and proposed a model to explain the memory device's behavior in such a design,[50] and below we give further details. Therefore, to prove that the coupling between the silicon back-gate and the FG through extra metal pad areas should not be negligible, we investigate its area impact on the device's memory window. For this, we designed dual-gate devices with different sizes of FG (here including the connection extra pad, see **Figure 1c**), on fully encapsulated hBN/SLG/hBN and capped hBN/SLG devices. **Figure 3a-b** shows the correlation between the position of the CNP and the extra pad sizes for several devices, demonstrating a clear enhancement in the memory window with larger pad sizes. The results also indicate a saturation effect for areas larger than 2500 μm$^2$. SM-Section 6 brings further discussion and optical images of devices measured with FG of distinct areas. Therefore, based on these results, we conclude that electrical hysteresis results from the capacitive coupling between the back and the top gates.

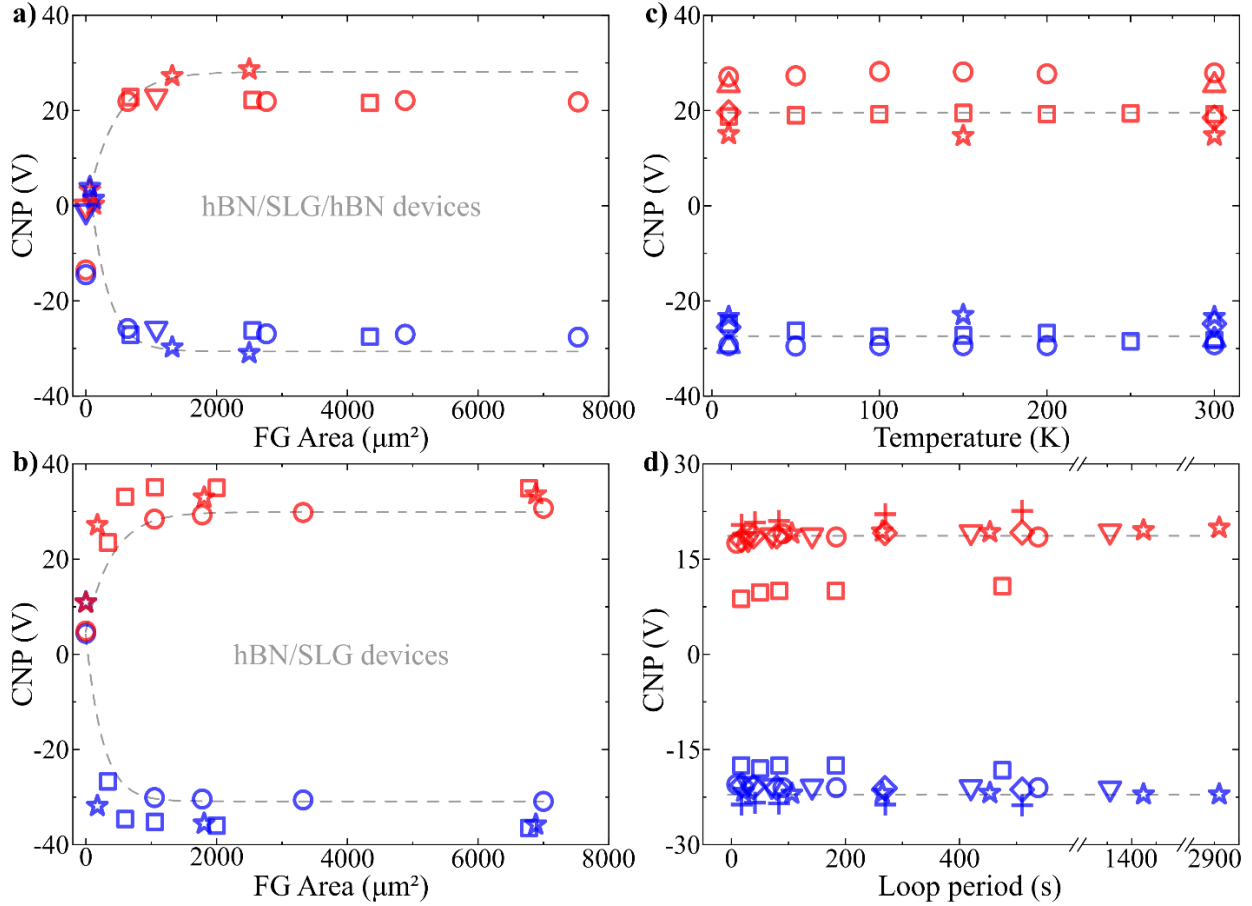

**Figure 3: a-b)** Memory window for GFETs with different FG areas for the encapsulated hBN/SLG/hBN and capped hBN/SLG devices, respectively. Symbols correspond to different devices, while the blue and red colors represent the forward and backward $V_{BG}$ sweeps, respectively. For all GFETs characterization, $|V_{BG, max}| = 40V$, vacuum, and room temperature. **c-d)** Position of the CNP for encapsulated GFETs in the forward (blue symbols) and backward (red symbols) $V_{BG}$ sweeps as a function of the temperature and $|V_{BG}|$ loop period, respectively. In (c), $|V_{BG, max}| = 40$ V while in (d), $|V_{BG, max}| = 30$ V to avoid leakage current for faster sweeps. Gray dashed lines serve as a guide to the eye. Test conditions are fixed with an extra metal pad size of $200 \times 200$ μm$^2$ (c-d), at 300 K (a,b,d), and in vacuum for all GFETs.

So far, we have shown that the memory window of non-volatile GFETs depends on both the top-hBN thickness and the area of the FG. Moreover, SM-Section 7 provides results demonstrating that these memory devices remain stable for over one year under laboratory conditions. The data reveals negligible changes over time, supporting the robustness of the encapsulation step for reliable memory device operation. It is also important to note that memory devices based on volatile processes exhibit a clear dependence on voltage bias sweep rate and operational temperature.[49–52] To further demonstrate that our GFETs exhibit a non-volatile response - one that should not depend on either parameter - we evaluated the memory performance at different $V_{BG}$ sweep rates and temperatures (ranging from 10 K to 300 K). **Figure 3c** plots the CNP for devices measured at different temperatures, and **Figure 3d** presents the CNP shift as a function of $|V_{BG}|$ loop period (sweep rate). The results indicate that the memory window is independent of both parameters. Moreover, it demonstrates storage capability and operation of these memory devices at cryogenic temperatures. Note that small variations for faster sweep rates can be associated with impurities, bubbles, or imperfections in the heterostructures or devices. SM-Section 8 provides the transfer curves of the GFETs under the various parameters analyzed.

To understand the charge transfer process and the physical origin of the electrical hysteresis observed in our GFETs, we consider that these non-volatile devices operate based on a charging mechanism driven by the displacement current induced by $V_{BG}$ sweeps. In this framework, the conductance is given by[50]

$$\sigma = \sigma_{res} + \frac{\mu n}{1+\tilde{\rho}\mu n}, \quad (1)$$

here $\mu$ is the carrier mobility in the graphene sheet, and the gate-independent resistive factor $\tilde{\rho}$ is defined as $\tilde{\rho} = \frac{1}{\sigma_{ADP}} + \rho_c$, where $\sigma_{ADP}$ represents the contribution from acoustic phonon scattering via the deformation potential, and $\rho_c$ is the parasitic contact resistance.

The carrier density depends on both the $V_{BG}$ and the charge fluctuation induced by the capacitive coupling with the FG, $\delta n$, according to $n = |C_g V_{BG} - \delta n|$. The charging and discharging of the FG are governed by a displacement current, which is driven by the sweep of the $V_{BG}$ and determined by an effective capacitance, $C_{eff}$

$$\frac{d\delta n}{dt} = C_{eff}\frac{dV_{BG}}{dt}, \qquad (2)$$

as described in the reference.[50] This effective capacitance, combined with the geometric capacitance $C_0$, forms the total gate capacitance $C = C_{eff}+C_0$, which controls the dynamic (displacement-driven) charging current

$$I = C\frac{dV_{BG}}{dt}. \qquad (3)$$

This mechanism differs fundamentally from volatile responses typically observed in GFETs, where the hysteresis arises from direct charging and discharging of the SLG channel through interface traps or dipolar adsorbates.[49,51,53] In those cases, the memory window is sensitive to both the sweep rate and environmental conditions, and the system tends to relax back to its equilibrium state once the perturbation is removed. In contrast, the behavior described here is governed by displacement current-driven charging of an FG, which is electrically isolated from the channel and acts as a long-term charge storage node. Because the charge is modulated indirectly via capacitive coupling, rather than direct injection, the system can exhibit a non-volatile response. This leads to the striking observation of hysteresis loops that are invariant with respect to the sweep period and persist without relaxation, highlighting the role of capacitive charge exchange in generating memory effects. The decoupling of dynamic response from the intrinsic relaxation time of the graphene sheet represents a distinctive and robust form of memory behavior that is not governed by conventional trap- or polarization-assisted hysteresis.

After understanding the physical mechanism that rules the operation of our memory device, we proceed with the performance characterization of this non-volatile system as proof of principle.

The basic operating principle of flash memory is to store electrons or holes in an FG layer; these stored charges shift the device's CNP. Here, the dynamic behavior of a representative GFET is investigated at 300 K and 10 K. Switching between different current states is achieved by applying $V_{BG, pulses} = \pm 20$ V, while the source electrode is grounded and the drain electrode is biased at $V_{SD} = 100$ mV. **Figure 4a** shows the $V_{BG, pulses}$ ($\pm 20$ V) applied over time to a GFET with an FG at 300 K, and **Figure 4b** displays the $I_{DS}$ current measured during the switching events. Initially, the device is in a high-current state ($I_{DS} \sim 48$ µA), corresponding to the "ON state". A positive $V_{BG, pulse} = +20$ V (100 ms) is then applied, leading to a ~90 µA current spike. When $V_{BG}$ returns to 0 V (2.4 s), the device switches to the "OFF" state, maintaining a stable low current of ~42 µA. The device is restored to the initial ON state by applying a symmetric negative $V_{BG, pulse} = -20$ V (100 ms), during which a ~65 µA current spike is observed. Overall, the GFET-based memory device can be switched reliably and stably between the "ON" and "OFF" states. The reliability of the memory device is crucial for non-volatile data storage; therefore, both retention time and cyclic endurance are investigated.

To assess the stability and tunability of the memory device, the dynamic behavior of the GFET with an FG is evaluated by applying $V_{BG,pulses} = \pm 20$ V for 100 ms at $V_{DS} = 100$ mV. **Figure 4c-d** shows the "ON" and "OFF" states collected from the GFET with FG at 300 K and 10 K, respectively. The device is programmed (erased) to the ON (OFF) state through $V_{BG, pulses}$ of $-20$ V ($+20$ V) with a pulse width of 100 ms. In both measurements, 9800 sequential voltage pulses were applied to the back gate while keeping the top-gate electrode disconnected (FG). These results demonstrate a clear distinction between the "ON" and "OFF" states during $V_{BG}$ cycles, exhibiting stable endurance characteristics, and confirm that memory devices based on GFETs with FG can operate at cryogenic temperatures.

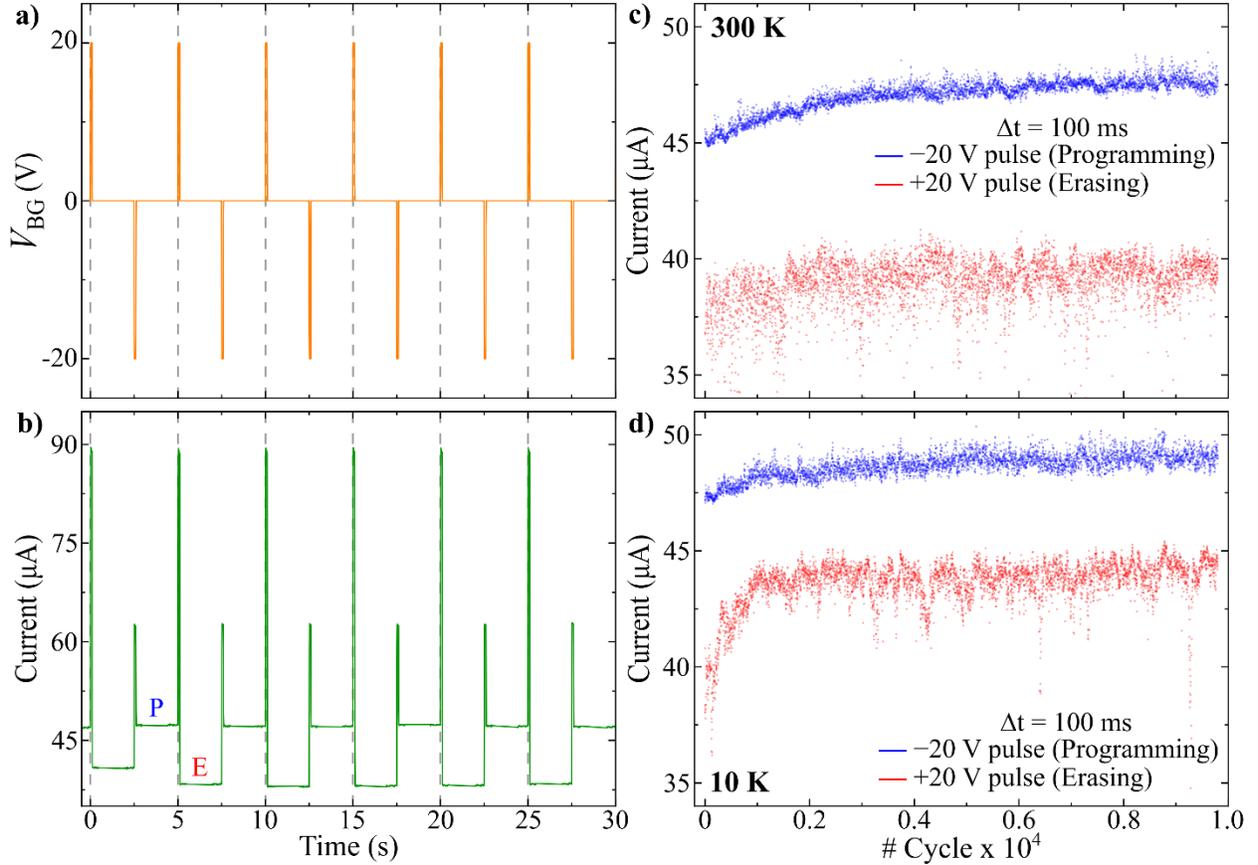

**Figure 4:** Memory states. **a-b)** The dynamic switching behavior between ON and OFF states by applying alternating $V_{BG}$ pulses (±20 V, 100 ms) with a time interval of 2.4 s. Each state was read at $V_{BG}$ = 0 V, $V_D$ = 100 mV after being programmed (erased) by one pulse voltage of +20 V (−20 V) and a width of Δt = 100 ms on the gate. **c-d)** Time evolution of current states after writing ($V_{BG,\,pulse}$ = −20 V, 100 ms) and erasing ($V_{BG,\,pulse}$ = +20 V, 100 ms) operations performed at 300 K and 10 K, respectively. The endurance test of the device was taken for 9800 cycles of the writing/erasing pulses at both temperatures by applying alternating $V_{BG}$ pulses (±20 V, 100 ms) with a time interval of 2.4 s. The current reads at $V_{BG}$ = 0 V, $V_{DS}$ = 100 mV, and vacuum. The extra metal pad size is 200 × 200 μm² in the GFETs.

Another important aspect of memory devices is the capability to reset the memory state at any time. **Figure 5a-b** shows the clear difference in memory retention time when the device is left with the top electrode (FG) either disconnected during measurements or grounded, respectively. While **Figure 5a** demonstrates that after 2 hours, the new state created by the +20V writing pulse for 50 s is distinct from the pre-stimulus state - behavior that is typical of non-volatile memory - **Figure 5b** shows that the device restores the old state as soon as $V_{BG}$ is set back to zero volts. This illustrates that by grounding the FG, we can reset any pre-written memory state, eliminating the excess charge present at the top electrode. Therefore, this extra probe works as an important control knob to restore the initial condition of the device.

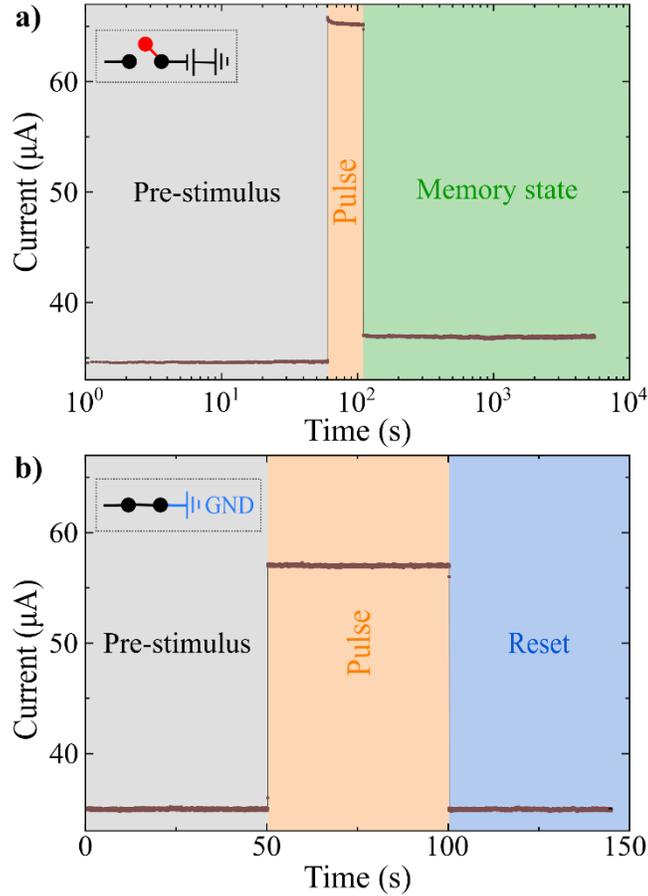

**Figure 5: a-b)** $I_{DS}$ response versus time showing a steady response after $V_{BG, pulse}$ = +20 V, $V_{DS}$ = 100 mV, and a reversible response when the top electrode is either floating or grounded (GND), respectively. The extra metal pad size is 200 × 200 μm² in the GFET, and the measurements were taken in vacuum at 300 K.

Although we present here a proof-of-principle demonstration of device operation, we foresee more elaborate studies where one can quantify the writing and erasing pulse widths and pulse amplitudes for multilevel storage applications. Moreover, the device design discussed here can be implemented beyond SLG and hBN heterostructures, due to the rich variety of 2D materials and stacking technologies, thus allowing the utilization of light to improve performance and functionality of optical memory devices.[54–57]

**CONCLUSIONS**

We have demonstrated a robust and versatile non-volatile memory device based on dual-gated hBN/graphene-based heterostructures, where a single metal electrode serves as a floating gate, top gate, or reset contact. Our findings reveal that the memory behavior arises from the capacitive coupling between the silicon back-gate and the floating gate through the hBN dielectric, enabling tunable and reversible charge storage. We then proposed a dynamic model that captures the essential physics of such device's response under gate modulation. The memory window is strongly dependent on the top-hBN thickness and floating gate area, reaching programming efficiencies above 90% and maintaining stable operation over long timescales and multiple switching cycles, even at cryogenic temperatures. The ability to electrically reset the device by grounding the top electrode further highlights its potential for reconfigurable and low-power memory architectures. This work provides a clear framework for engineering scalable 2D memory systems and paves the way toward their integration in high-density and quantum-compatible electronics.

**METHODS**

**Sample fabrication** – All samples were fabricated and characterized by Raman and Atomic Force Microscope (AFM) after stacking the individual layers by a dry transfer technique[58]. Following the heterostructure assembly, device fabrication was carried out using EBL and reactive ion etching (RIE) plasma to define the contact regions and device geometry, followed by standard nanofabrication procedures. The e-beam acceleration used in the EBL step for the device fabrication was 20kV, except for the data in SM-Section 5. See SM-Section 1 for more experimental details.

**Device electrical characterization** - All GFETs were characterized in a Lakeshore CPX-EM-HF cryogen-free closed cycle refrigerator probe station under a pressure of ~$10^{-6}$ Torr. The GFETs were measured as follows: transfer curves under back and top-gates were obtained in a two-terminal configuration using standard DC techniques with a Keithley 4200A-SCS Parameter Analyzer and $V_{DS}$ = 100 mV. For hysteresis loops, the $V_{BG}$ ($V_{TG}$) bias was swept continuously from zero volts to the maximum negative $V_{BG}$ ($V_{TG}$), then to the maximum positive $V_{BG}$ ($V_{TG}$), back to the negative $V_{BG}$ ($V_{TG}$), and finally returned to zero volts. Data acquisition was performed throughout the entire voltage sweep. The pulsed data collection and time-dependent response under bias stimuli were performed in a Keithley 2636A source measure unit controlled by a homemade LabVIEW program.

**Author Contributions**

A.R.C. designed and coordinated the study. G.R.L., A.B.Y., and A.R.C. fabricated, characterized the devices, and analyzed the experimental data. G.S.S. and N.T.K.B.J. developed the LabView program and performed the pulsed characterization together with G.R.L. under the supervision of R.F.O. and A.R.C. Theoretical modelling was developed by I.R. and V.L-R., while K.W. and T.T. supplied the hBN crystals. The manuscript was written through the contributions of all authors. All authors have approved the final version of the manuscript.

**ACKNOWLEDGMENT**

The authors acknowledge financial support from the Brazilian Ministry of Science, Technology and Innovation (MCTI); the National Council for Scientific and Technological Development (CNPq, grant numbers 309920/2021-3 and 311536/2022-0); the Coordination for the Improvement of Higher Education Personnel (CAPES); and the Sao Paulo Research Foundation (FAPESP, grant numbers66 2019/14949-9 and 2023/09395-0). The authors also acknowledge the Brazilian Nanotechnology National Laboratory (LNNano) and Brazilian Synchrotron Light Laboratory (LNLS), both part of the Brazilian Centre for Research in Energy and Materials (CNPEM), a private nonprofit organization under the supervision of the Brazilian Ministry for Science, Technology, and Innovations (MCTI), for device fabrication and characterization - LNNano/CNPEM (proposals: MNF-20240039, MNF- 20240040, MNF-20250043, and MNF-20250045) and LAM-2D (proposal: 20240497) at LNLS/CNPEM. The Atomic Force Microscopy Laboratory staff at LNNano/CNPEM is acknowledged for their assistance (MFA-20240042 and MFA-20250047), besides Marcelo R. Piton, Davi H.S. Camargo, and Leirson D. Palermo for the experimental assistance. K.W. and T.T. acknowledge support from the JSPS KAKENHI (grant numbers 21H05233 and 23H02052) and World Premier International Research Center Initiative (WPI), MEXT, Japan. Finally, A.R.C. acknowledges Prof. Andres Castellanos-Gomez and Mr. Thomas Pucher for enlightening discussions during the 38th IWEPNM, Kirchberg/Tirol, Austria.

# REFERENCES

(1) Hwang, C. S. Prospective of Semiconductor Memory Devices: From Memory System to Materials. *Adv Electron Mater* **2015**, *1* (6). https://doi.org/10.1002/aelm.201400056.

(2) *International Roadmap for Devices and Systems (IRDS*. https://irds.ieee.org/.

(3) Wu, L.; Wang, A.; Shi, J.; Yan, J.; Zhou, Z.; Bian, C.; Ma, J.; Ma, R.; Liu, H.; Chen, J.; Huang, Y.; Zhou, W.; Bao, L.; Ouyang, M.; Pennycook, S. J.; Pantelides, S. T.; Gao, H.-J. Atomically Sharp Interface Enabled Ultrahigh-Speed Non-Volatile Memory Devices. *Nat Nanotechnol* **2021**, *16* (8), 882–887. https://doi.org/10.1038/s41565-021-00904-5.

(4) Kim, S. S.; Yong, S. K.; Kim, W.; Kang, S.; Park, H. W.; Yoon, K. J.; Sheen, D. S.; Lee, S.; Hwang, C. S. Review of Semiconductor Flash Memory Devices for Material and Process Issues. *Advanced Materials* **2023**, *35* (43). https://doi.org/10.1002/adma.202200659.

(5) Hu, C.; Liang, L.; Yu, J.; Cheng, L.; Zhang, N.; Wang, Y.; Wei, Y.; Fu, Y.; Wang, Z. L.; Sun, Q. Neuromorphic Floating-Gate Memory Based on 2D Materials. *Cyborg and Bionic Systems* **2025**, *6*. https://doi.org/10.34133/cbsystems.0256.

(6) Zhang, Q.; Zhang, Z.; Li, C.; Xu, R.; Yang, D.; Sun, L. Van Der Waals Materials-Based Floating Gate Memory for Neuromorphic Computing. *Chip* **2023**, *2* (4), 100059. https://doi.org/10.1016/j.chip.2023.100059.

(7) Choi, M. S.; Lee, G.-H.; Yu, Y.-J.; Lee, D.-Y.; Lee, S. H.; Kim, P.; Hone, J.; Yoo, W. J. Controlled Charge Trapping by Molybdenum Disulphide and Graphene in Ultrathin Heterostructured Memory Devices. *Nat Commun* **2013**, *4*, 1624. https://doi.org/10.1038/ncomms2652.

(8) Wang, H.; Guo, H.; Guzman, R.; JiaziLa, N.; Wu, K.; Wang, A.; Liu, X.; Liu, L.; Wu, L.; Chen, J.; Huan, Q.; Zhou, W.; Yang, H.; Pantelides, S. T.; Bao, L.; Gao, H. Ultrafast Non-Volatile Floating-Gate Memory Based on All-2D Materials. *Advanced Materials* **2024**, *36* (24). https://doi.org/10.1002/adma.202311652.

(9) Mallik, S. K.; Padhan, R.; Sahu, M. C.; Pradhan, G. K.; Sahoo, P. K.; Dash, S. P.; Sahoo, S. Ionotronic WS2 Memtransistors for 6-Bit Storage and Neuromorphic Adaptation at High Temperature. *NPJ 2D Mater Appl* **2023**, *7* (1), 63. https://doi.org/10.1038/s41699-023-00427-8.

(10) Gadelha, A. C.; Cadore, A. R.; Watanabe, K.; Taniguchi, T.; de Paula, A. M.; Malard, L. M.; Lacerda, R. G.; Campos, L. C. Gate-Tunable Non-Volatile Photomemory Effect in MoS2 Transistors. *2d Mater* **2019**, *6* (2), 025036. https://doi.org/10.1088/2053-1583/ab0af1.

(11) Wang, S.; He, C.; Tang, J.; Lu, X.; Shen, C.; Yu, H.; Du, L.; Li, J.; Yang, R.; Shi, D.; Zhang, G. New Floating Gate Memory with Excellent Retention Characteristics. *Adv Electron Mater* **2019**, *5* (4). https://doi.org/10.1002/aelm.201800726.

(12) Bertolazzi, S.; Krasnozhon, D.; Kis, A. Nonvolatile Memory Cells Based on MoS 2 /Graphene Heterostructures. *ACS Nano* **2013**, *7* (4), 3246–3252. https://doi.org/10.1021/nn3059136.

(13) Liu, C.; Pan, J.; Yuan, Q.; Zhu, C.; Liu, J.; Ge, F.; Zhu, J.; Xie, H.; Zhou, D.; Zhang, Z.; Zhao, P.; Tian, B.; Huang, W.; Wang, L. Highly Reliable Van Der Waals Memory Boosted by a Single 2D Charge Trap Medium. *Advanced Materials* **2024**, *36* (3). https://doi.org/10.1002/adma.202305580.

(14) Kim, S. H.; Yi, S.-G.; Park, M. U.; Lee, C.; Kim, M.; Yoo, K.-H. Multilevel MoS2 Optical Memory with Photoresponsive Top Floating Gates. *ACS Appl Mater Interfaces* **2019**, *11* (28), 25306–25312. https://doi.org/10.1021/acsami.9b05491.

(15) Dastgeer, G.; Nisar, S.; Rasheed, A.; Akbar, K.; Chavan, V. D.; Kim, D.; Wabaidur, S. M.; Zulfiqar, M. W.; Eom, J. Atomically Engineered, High-Speed Non-Volatile Flash Memory Device Exhibiting Multibit Data Storage Operations. *Nano Energy* **2024**, *119*, 109106. https://doi.org/10.1016/j.nanoen.2023.109106.

(16) Cao, W.; Bu, H.; Vinet, M.; Cao, M.; Takagi, S.; Hwang, S.; Ghani, T.; Banerjee, K. The Future Transistors. *Nature* **2023**, *620* (7974), 501–515. https://doi.org/10.1038/s41586-023-06145-x.

(17) Chhowalla, M.; Jena, D.; Zhang, H. Two-Dimensional Semiconductors for Transistors. *Nat Rev Mater* **2016**, *1* (11), 16052. https://doi.org/10.1038/natrevmats.2016.52.

(18) Liu, L.; Liu, C.; Jiang, L.; Li, J.; Ding, Y.; Wang, S.; Jiang, Y.-G.; Sun, Y.-B.; Wang, J.; Chen, S.; Zhang, D. W.; Zhou, P. Ultrafast Non-Volatile Flash Memory Based on van Der Waals Heterostructures. *Nat Nanotechnol* **2021**, *16* (8), 874–881. https://doi.org/10.1038/s41565-021-00921-4.

(19) Pervez, M. H.; Elahi, E.; Khan, M. A.; Nasim, M.; Asim, M.; Rehmat, A.; Rehman, M. A.; Assiri, M. A.; Rehman, S.; Eom, J.; Khan, M. F. Recent Developments on Novel 2D Materials for Emerging Neuromorphic Computing Devices. *Small Struct* **2025**, *6* (2). https://doi.org/10.1002/sstr.202400386.

(20) Zhang, Y.; Huang, C.-H.; Nomura, K. Dual-Gated Ambipolar Oxide Synaptic Transistor for Multistate Excitatory and Inhibitory Responses. *Appl Phys Lett* **2022**, *121* (26). https://doi.org/10.1063/5.0123309.

(21) Selmi, G. S.; Lourenço Neto, E. R.; Lelis, G. C.; Okazaki, A. K.; Riul, A.; Braunger, M. L.; de Oliveira, R. F. Pulse Dynamics in Reduced Graphene Oxide Electrolyte-Gated Transistors: Charge Memory Effects and Mechanisms Governing the Ion-To-Electron Transduction. *Adv Electron Mater* **2025**, *11* (8). https://doi.org/10.1002/aelm.202400791.

(22) Ames, A.; Sousa, F. B.; Souza, G. A. D.; de Oliveira, R.; Silva, I. R. F.; Rodrigues, G. L.; Watanabe, K.; Taniguchi, T.; Marques, G. E.; Barcelos, I. D.; Cadore, A. R.; Lopez-Richard, V.; Teodoro, M. D. Optical Memory in a MoSe2/Clinochlore Device. *ACS Appl Mater Interfaces* **2025**, *17* (8), 12818–12826. https://doi.org/10.1021/acsami.4c19337.

(23) Lopez-Richard, V.; Filgueira e Silva, I. R.; Ames, A.; Sousa, F. B.; Teodoro, M. D.; Barcelos, I. D.; de Oliveira, R.; Cadore, A. R. The Emergence of Mem-Emitters. *Nano Lett* **2024**. https://doi.org/10.1021/acs.nanolett.4c04586.

(24) Sadaf, M. U. K.; Chen, Z.; Subbulakshmi Radhakrishnan, S.; Sun, Y.; Ding, L.; Graves, A. R.; Yang, Y.; Redwing, J. M.; Das, S. Enabling Static Random-Access Memory Cell Scaling with Monolithic 3D Integration of 2D Field-Effect Transistors. *Nat Commun* **2025**, *16* (1), 4879. https://doi.org/10.1038/s41467-025-59993-8.

(25) Zhang, N.; Zhang, Z.; Feng, R.; Chen, Y.; Li, Q.; Yang, Y.; Jiang, M.; Zhao, W.; Zhu, Z.; Zhou, X.; Li, Z. Gate-Tunable Polarization Gradient in 2D Polar Semiconductor for Synaptic Transistor. *ACS Nano* **2025**. https://doi.org/10.1021/acsnano.5c07480.

(26) Rios, A. C.; Aarão-Rodrigues, L.; Cadore, A. R.; de Andrade, R. R.; Montoro, L. A.; Malachias, A. Tailoring Resistive Switching Properties of TiO2 with Controlled Incorporation of Oxide Nanoparticles. *Mater Res Express* **2016**, *3* (8), 085024. https://doi.org/10.1088/2053-1591/3/8/085024.

(27) Hong, A. J.; Song, E. B.; Yu, H. S.; Allen, M. J.; Kim, J.; Fowler, J. D.; Wassei, J. K.; Park, Y.; Wang, Y.; Zou, J.; Kaner, R. B.; Weiller, B. H.; Wang, K. L. Graphene Flash Memory. *ACS Nano* **2011**, *5* (10), 7812–7817. https://doi.org/10.1021/nn201809k.

(28) Ding, Y.; Liu, L.; Li, J.; Cao, R.; Jiang, Y.; Liu, C.; Liu, Q.; Zhou, P. A Semi-Floating Memory with 535% Enhancement of Refresh Time by Local Field Modulation. *Adv Funct Mater* **2020**, *30* (15). https://doi.org/10.1002/adfm.201908089.

(29) He, C.; Tang, J.; Shang, D.-S.; Tang, J.; Xi, Y.; Wang, S.; Li, N.; Zhang, Q.; Lu, J.-K.; Wei, Z.; Wang, Q.; Shen, C.; Li, J.; Shen, S.; Shen, J.; Yang, R.; Shi, D.; Wu, H.; Wang, S.; Zhang, G. Artificial Synapse Based on van Der Waals Heterostructures with Tunable Synaptic Functions for Neuromorphic Computing. *ACS Appl Mater Interfaces* **2020**, *12* (10), 11945–11954. https://doi.org/10.1021/acsami.9b21747.

(30) Sasaki, T.; Ueno, K.; Taniguchi, T.; Watanabe, K.; Nishimura, T.; Nagashio, K. Ultrafast Operation of 2D Heterostructured Nonvolatile Memory Devices Provided by the Strong Short-Time Dielectric Breakdown Strength of *h*-BN. *ACS Appl Mater Interfaces* **2022**, *14* (22), 25659–25669. https://doi.org/10.1021/acsami.2c03198.

(31) Shi, J.; Liu, Z.; Wei, J.; Azam, A.; Lin, C.-H.; Liu, Y.; Li, S. Revolutionizing Nonvolatile Memory: Advances and Future Prospects of 2D Floating-Gate Technology. *ACS Nano* **2025**. https://doi.org/10.1021/acsnano.5c02740.

(32) Zhang, Z.; Yang, D.; Li, H.; Li, C.; Wang, Z.; Sun, L.; Yang, H. 2D Materials and van Der Waals Heterojunctions for Neuromorphic Computing. *Neuromorphic Computing and Engineering* **2022**, *2* (3), 032004. https://doi.org/10.1088/2634-4386/ac8a6a.

(33) Liu, C.; Chen, H.; Wang, S.; Liu, Q.; Jiang, Y.-G.; Zhang, D. W.; Liu, M.; Zhou, P. Two-Dimensional Materials for next-Generation Computing Technologies. *Nat Nanotechnol* **2020**, *15* (7), 545–557. https://doi.org/10.1038/s41565-020-0724-3.

(34) Du, X.; Skachko, I.; Barker, A.; Andrei, E. Y. Approaching Ballistic Transport in Suspended Graphene. *Nat Nanotechnol* **2008**, *3* (8), 491–495. https://doi.org/10.1038/nnano.2008.199.

(35) Ranjan, A.; Raghavan, N.; Holwill, M.; Watanabe, K.; Taniguchi, T.; Novoselov, K. S.; Pey, K. L.; O'Shea, S. J. Dielectric Breakdown in Single-Crystal Hexagonal Boron Nitride. *ACS Appl Electron Mater* **2021**, *3* (8), 3547–3554. https://doi.org/10.1021/acsaelm.1c00469.

(36) Yankowitz, M.; Ma, Q.; Jarillo-Herrero, P.; LeRoy, B. J. Van Der Waals Heterostructures Combining Graphene and Hexagonal Boron Nitride. *Nature Reviews Physics* **2019**, *1* (2), 112–125. https://doi.org/10.1038/s42254-018-0016-0.

(37) Watanabe, K.; Taniguchi, T.; Kanda, H. Direct-Bandgap Properties and Evidence for Ultraviolet Lasing of Hexagonal Boron Nitride Single Crystal. *Nat Mater* **2004**, *3* (6), 404–409.

(38) Xue, J.; Sanchez-Yamagishi, J.; Bulmash, D.; Jacquod, P.; Deshpande, A.; Watanabe, K.; Taniguchi, T.; Jarillo-Herrero, P.; LeRoy, B. J. Scanning Tunnelling Microscopy and Spectroscopy of Ultra-Flat Graphene on Hexagonal Boron Nitride. *Nat Mater* **2011**, *10* (4), 282–285. https://doi.org/10.1038/nmat2968.

(39) Purdie, D. G.; Pugno, N. M.; Taniguchi, T.; Watanabe, K.; Ferrari, A. C.; Lombardo, A. Cleaning Interfaces in Layered Materials Heterostructures. *Nat Commun* **2018**, *9* (1), 1–12. https://doi.org/10.1038/s41467-018-07558-3.

(40) Viti, L.; Cadore, A. R.; Yang, X.; Vorobiev, A.; Muench, J. E.; Watanabe, K.; Taniguchi, T.; Stake, J.; Ferrari, A. C.; Vitiello, M. S. Thermoelectric Graphene Photodetectors with Sub-Nanosecond Response Times at Terahertz Frequencies. *Nanophotonics* **2020**, *10* (1), 89–98. https://doi.org/10.1515/nanoph-2020-0255.

(41) Wang, J.; Ma, F.; Sun, M. Graphene, Hexagonal Boron Nitride, and Their Heterostructures: Properties and Applications. *RSC Adv.* **2017**, *7* (27), 16801–16822. https://doi.org/10.1039/C7RA00260B.

(42) Dean, C. R.; Young, A. F.; Meric, I.; Lee, C.; Wang, L.; Sorgenfrei, S.; Watanabe, K.; Taniguchi, T.; Kim, P.; Shepard, K. L.; Hone, J. Boron Nitride Substrates for High-Quality Graphene Electronics. *Nat Nanotechnol* **2010**, *5* (10), 722–726. https://doi.org/10.1038/nnano.2010.172.

(43) Laturia, A.; Van de Put, M. L.; Vandenberghe, W. G. Dielectric Properties of Hexagonal Boron Nitride and Transition Metal Dichalcogenides: From Monolayer to Bulk. *NPJ 2D Mater Appl* **2018**, *2* (1), 6. https://doi.org/10.1038/s41699-018-0050-x.

(44) Hattori, Y.; Taniguchi, T.; Watanabe, K.; Nagashio, K. Layer-by-Layer Dielectric Breakdown of Hexagonal Boron Nitride. *ACS Nano* **2015**, *9* (1), 916–921. https://doi.org/10.1021/nn506645q.

(45) Mania, E.; Cadore, A. R.; Taniguchi, T.; Watanabe, K.; Campos, L. C. Topological Valley Transport at the Curved Boundary of a Folded Bilayer Graphene. *Commun Phys* **2019**, *2* (1), 2–7. https://doi.org/10.1038/s42005-018-0106-4.

(46) Mania, E.; Alencar, A. B.; Cadore, A. R.; Carvalho, B. R.; Watanabe, K.; Taniguchi, T.; Neves, B. R. A.; Chacham, H.; Campos, L. C. Spontaneous Doping on High Quality Talc-Graphene-HBN van Der Waals Heterostructures. *2d Mater* **2017**, *4* (3), 031008. https://doi.org/10.1088/2053-1583/aa76f4.

(47) Yan, J.; Fuhrer, M. S. Charge Transport in Dual Gated Bilayer Graphene with Corbino Geometry. *Nano Lett* **2010**, *10* (11), 4521–4525. https://doi.org/10.1021/nl102459t.

(48) Shi, W.; Kahn, S.; Jiang, L.; Wang, S. Y.; Tsai, H. Z.; Wong, D.; Taniguchi, T.; Watanabe, K.; Wang, F.; Crommie, M. F.; Zettl, A. Reversible Writing of High-Mobility and High-Carrier-Density Doping Patterns in Two-Dimensional van Der Waals Heterostructures. *Nat Electron* **2020**, *3* (2), 99–105. https://doi.org/10.1038/s41928-019-0351-x.

(49) Cadore, A. R.; Mania, E.; Watanabe, K.; Taniguchi, T.; Lacerda, R. G.; Campos, L. C. Thermally Activated Hysteresis in High Quality Graphene/h-BN Devices. *Appl Phys Lett* **2016**, *108* (23), 233101. https://doi.org/10.1063/1.4953162.

(50) Lopez-Richard, V.; Silva, I. R. F. e; Rodrigues, G. L.; Watanabe, K.; Taniguchi, T.; Cadore, A. R. Frequency as a Clock: Synchronization and Intrinsic Recovery in Graphene Transistor Dynamics. **2025**.

(51) Wang, H.; Wu, Y.; Cong, C.; Shang, J.; Yu, T. Hysteresis of Electronic Transport in Graphene Transistors. *ACS Nano* **2010**, *4* (12), 7221–7228. https://doi.org/10.1021/nn101950n.

(52) Sahoo, A.; Nafday, D.; Paul, T.; Ruiter, R.; Roy, A.; Mostovoy, M.; Banerjee, T.; Saha-Dasgupta, T.; Ghosh, A. Out-of-Plane Interface Dipoles and Anti-Hysteresis in Graphene-Strontium Titanate Hybrid Transistor. *NPJ 2D Mater Appl* **2018**, *2* (1), 9. https://doi.org/10.1038/s41699-018-0055-5.

(53) Kurchak, A. I.; Morozovska, A. N.; Strikha, M. V. Hysteretic Phenomena in GFET: Comprehensive Theory and Experiment. *J Appl Phys* **2017**, *122* (4). https://doi.org/10.1063/1.4996095.

(54) Lee, D.; Hwang, E.; Lee, Y.; Choi, Y.; Kim, J. S.; Lee, S.; Cho, J. H. Multibit MoS2 Photoelectronic Memory with Ultrahigh Sensitivity. *Advanced Materials* **2016**, *28* (41), 9196–9202. https://doi.org/10.1002/adma.201603571.

(55) Vu, Q. A.; Lee, J. H.; Nguyen, V. L.; Shin, Y. S.; Lim, S. C.; Lee, K.; Heo, J.; Park, S.; Kim, K.; Lee, Y. H.; Yu, W. J. Tuning Carrier Tunneling in van Der Waals Heterostructures for Ultrahigh Detectivity. *Nano Lett* **2017**, *17* (1), 453–459. https://doi.org/10.1021/acs.nanolett.6b04449.

(56) Chen, Y.; Yu, J.; Zhuge, F.; He, Y.; Zhang, Q.; Yu, S.; Liu, K.; Li, L.; Ma, Y.; Zhai, T. An Asymmetric Hot Carrier Tunneling van Der Waals Heterostructure for Multibit Optoelectronic Memory. *Mater Horiz* **2020**, *7* (5), 1331–1340. https://doi.org/10.1039/C9MH01923E.

(57) Tran, M. D.; Kim, H.; Kim, J. S.; Doan, M. H.; Chau, T. K.; Vu, Q. A.; Kim, J.; Lee, Y. H. Two-Terminal Multibit Optical Memory via van Der Waals Heterostructure. *Advanced Materials* **2019**, *31* (7). https://doi.org/10.1002/adma.201807075.

(58) Cadore, A. R.; Rosa, B. L. T.; Paradisanos, I.; Mignuzzi, S.; De Fazio, D.; Alexeev, E. M.; Dagkli, A.; Muench, J. E.; Kakavelakis, G.; Shinde, S. M.; Yoon, D.; Tongay, S.; Watanabe, K.; Taniguchi, T.; Lidorikis, E.; Goykhman, I.; Soavi, G.; Ferrari, A. C. Monolayer WS2 Electro- and Photo-Luminescence Enhancement by TFSI Treatment. *2d Mater* **2024**, *11* (2), 025017. https://doi.org/10.1088/2053-1583/ad1a6a.

# Supporting Information:

# Graphene Heterostructure-Based Non-Volatile Memory Devices with Top Floating Gate Programming

*Gabriel L. Rodrigues[1,2], Ana B. Yoshida[1,2], Guilherme S. Selmi[1,2], Nickolas T.K.B de Jesus[1], Igor Ricardo[3], Kenji Watanabe[4], Takashi Taniguchi[5], Rafael F. de Oliveira[1], Victor Lopez-Richard[3], Alisson R. Cadore[1,6],\**

[1]Brazilian Nanotechnology National Laboratory (LNNano), Brazilian Center for Research in Energy and Materials (CNPEM), 13083-200, Campinas, Sao Paulo, Brazil
[2]"Gleb Wataghin" Institute of Physics, State University of Campinas, 13083-970, Campinas, São Paulo, Brazil
[3]Departamento de Física, Universidade Federal de São Carlos, 13565-905, São Carlos, São Paulo, Brazil
[4]Research Center for Electronic and Optical Materials, National Institute for Materials Science, 1-1 Namiki, 305-0044, Tsukuba, Japan
[5]Research Center for Materials Nanoarchitectonics, National Institute for Materials Science, 1-1 Namiki, 305-0044, Tsukuba, Japan
[6]Programa de Pós-Graduação em Física, Instituto de Física, Universidade Federal de Mato Grosso, 78060-900, Cuiabá, Mato Grosso, Brazil

*alisson.cadore@lnnano.cnpem.br

## Material Content:

**S.1 – Sample and Device Fabrication**

**S.2 – Electrical Measurements in Fully Encapsulated Devices with Grounded FG**

**S.3 – Reversibility and Gate Loop Direction Invariance**

**S.4 – FG Effect in hBN/SLG Devices**

**S.5 – EBL Acceleration Independence**

**S.6 – Extra Metal Pads Influence**

**S.7 – Stability of Memory Devices Over 1 Year**

**S.8 – Electrical Characteristics at Cryogenic Temperatures and Different Sweep Rates**

**S.9 – Performance Comparison**

## S.1 – Sample and Device Fabrication

High-quality exfoliated flakes of hBN, few-layer graphene (FLG), and single-layer graphene (SLG) were prepared via mechanical exfoliation onto separate SiO$_2$/Si substrates with a 285 nm thermally grown SiO$_2$ layer. The thickness of the selected top- and bottom-hBN flakes was initially determined by optical contrast, followed by atomic force microscopy (AFM) measurements after the heterostructure stacking. SLG flakes were identified by optical microscopy and confirmed using Raman spectroscopy.

The heterostructures were assembled using a dry transfer technique based on a polycarbonate (PC)/polydimethylsiloxane (PDMS) stamp.[1] Initially, the top-hBN flake was picked up with the PC/PDMS stamp at ~100 °C. Subsequently, two FLG flakes were aligned and picked up to work as source (S) and drain (D) electrodes, followed by the SLG and bottom-hBN flake at the same temperature. The complete stack (hBN/FLG-SLG-FLG/hBN) was then released onto a clean SiO$_2$/Si substrate by heating the stage to ~180 °C. The PC film was removed by soaking the sample in chloroform for several hours, followed by rinsing in ethanol and drying with nitrogen gas.

Following the heterostructure assembly, the device fabrication was carried out using electron beam lithography (EBL) to define the contact regions and device geometry. The contact regions (S, D, and FG electrodes) were patterned using EBL with an acceleration voltage of 5 - 30 kV and 300 μC/cm$^2$ as electron dose. Metal contacts (5/50 nm, Ti/Au) were subsequently deposited via electron beam evaporation, followed by a standard lift-off process. **Figure S1a** shows optical images of a representative hBN/FLG-SLG-FLG/hBN GFET fabricated and measured before and after FG fabrication.

Samples without FLG flakes were also fabricated following a similar procedure. For these samples, reactive ion etching (RIE) was employed to selectively etch the exposed top-hBN (SF$_6$ gas) and SLG (O$_2$ gas) areas to define the final GFET geometry. **Figure S1b** shows an optical image of a representative hBN/SLG/hBN GFET fabricated with side contacts and with FG electrodes of different areas.

Finally, devices based on a capped top-hBN/SLG stack were fabricated as follows. i) SLG flakes were exfoliated onto SiO$_2$/Si. ii) O$_2$ plasma etching was used to define the width of the SLG channel. iii) S and D electrodes were fabricated using EBL, metallization, and lift-off processes.

iv) Selected top-hBN flakes were picked up from SiO$_2$/Si substrates and transferred onto the desired areas. v) Finally, the top FG electrodes were fabricated in a subsequent step. **Figure S1c** shows an optical image of a representative capped hBN/SLG GFET fabricated with FG electrodes with different areas.

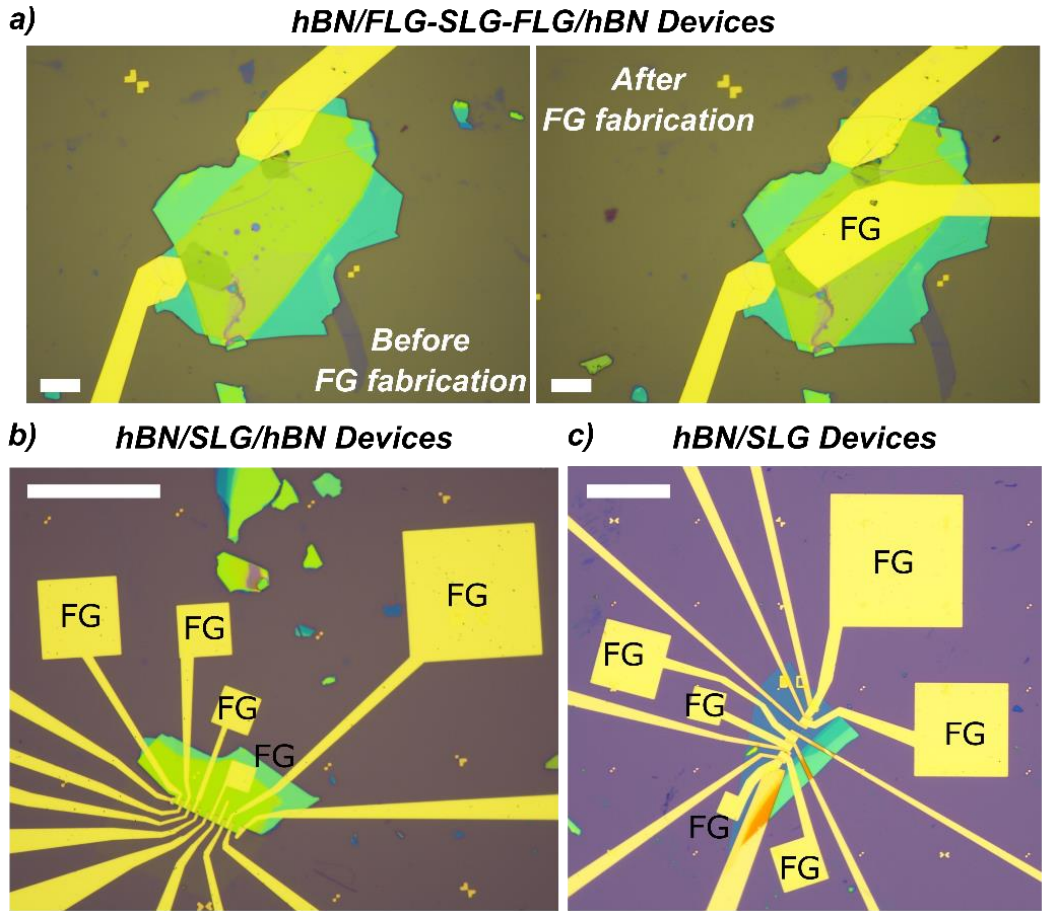

**Figure S1.** Optical images of GFETs fabricated: **a)** hBN/FLG-SLG-FLG/hBN device with and without FG fabrication. **b**) hBN/SLG/hBN devices with FG electrodes of different extra metal pad areas. **c)** hBN/SLG devices with FG electrodes of different areas. In all images, the FG electrodes are indicated, and all other electrodes are source and drain for each defined device. The scale bar represents 10 µm in (a), and 50 µm in (b-c).

## S.2 – Electrical Measurements in Fully Encapsulated Devices with Grounded FG

**Figure S2a-c** shows electrical measurements of a representative GFET taken at different bias conditions. The results indicate robust electrical hysteresis for $V_{BG}$ sweeps with the FG disconnected while it shows no hysteresis when it is set as $V_{TG} = 0$ V. The device shows negligible hysteresis under $V_{TG}$ sweeps, as expected.

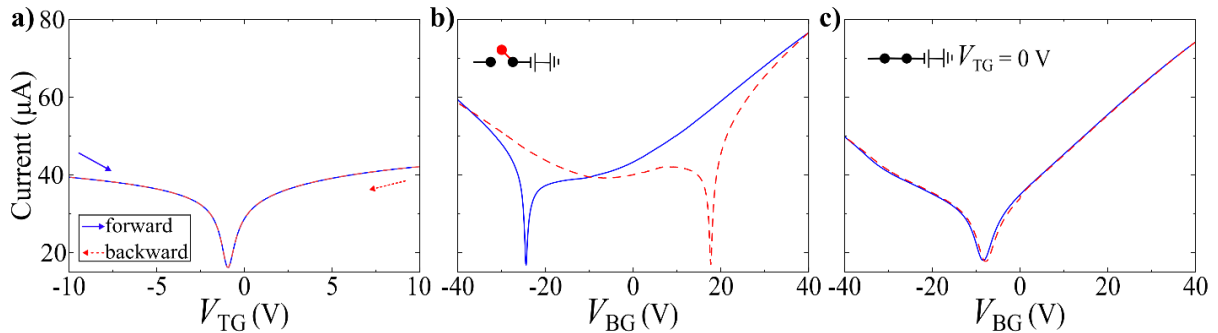

**Figure S2.** Transfer characteristics of the device operated in a **(a)** top-gate configuration; and **(b-c)** controlled by the back-gate, while the top electrode is floating or set as $V_{TG} = 0$ V, respectively. Forward (solid blue line) and backward (dashed red line) indicate the direction of the bias sweep applied. Test conditions are fixed $V_{DS} = 100$ mV and in vacuum at 300 K.

## S.3 – Reversibility and Gate Loop Direction Invariance

**Figure 3a-b** shows sequential electrical measurements increasing (Up) and decreasing (Down) the maximum $|V_{BG}|$ applied at the same sweep conditions, respectively. The results extend the consistency of hysteresis by being independent of the direction of increase (decrease) of the applied $|V_{BG}|$, but totally dependent on its amplitude. Furthermore, this demonstrates the reversibility of charge programming, in which electrical hysteresis is decreased and even deactivated with a reduction in $|V_{BG}|$. **Figure 3c** shows transfer curves in different sweep directions, in which one can observe the invariance of the gate loop course for the memory effect.

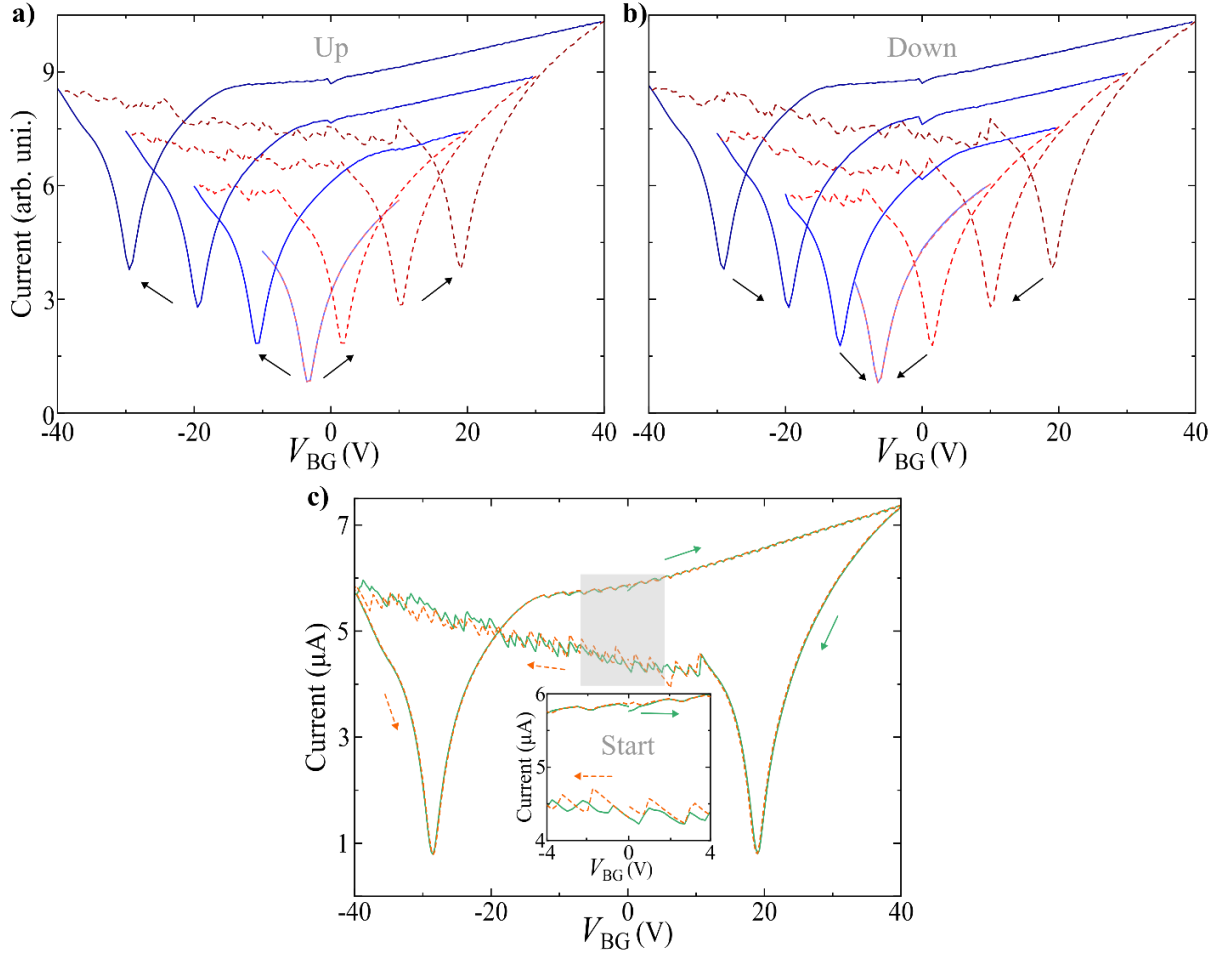

**Figure S3. a-b)** Transfer curves of a GFET with the increasing (Up) and the decreasing of the maximum |$V_{BG}$|, while the top electrode is floating, respectively. **c)** Transfer curves measured for different $V_{BG}$ loop directions: 0 V → +40 V → −40 V → 0 V (solid green line) and 0 V → −40 V → +40 V → 0 V (dashed orange line). The black arrows indicate the increase (a) and decrease (b) of the maximum $V_{BG}$, while the green and orange arrows in (c) are related to the start of each loop direction. Test conditions are fixed $V_{DS}$ =10 mV, extra metal pad size is 200 × 200 μm², and in vacuum at 300 K.

### S.4 – FG Effect in Capped hBN/SLG Devices

**Figure S4a-b** shows the schematic structure and electrical connections of the fabricated dual-gated memory device based on a capped hBN/SLG heterostructures. The source-drain current ($I_{DS}$) in such GFETs can be modulated by the $V_{BG}$, $V_{TG}$, and FG voltages. **Figure S4c** shows an optical image of a representative GFET fabricated with and without FG electrodes. Dev-1 and Dev-2 devices share the same SLG and hBN layers. They share one electrode as the drain but differ by the presence of the FG electrode: Dev-1 is only back-gated, while Dev-2 is dual-gated. The transfer curves obtained from the individual devices are shown in **Figure S4d-f** and display a rather different electrical behavior. While the device without FG shows no electrical hysteresis, the one with FG confirms the existence of a very large split in the CNP position for the $V_{BG}$ loop. Finally, Dev-2 shows no electrical hysteresis for $V_{TG}$ loops, as expected.

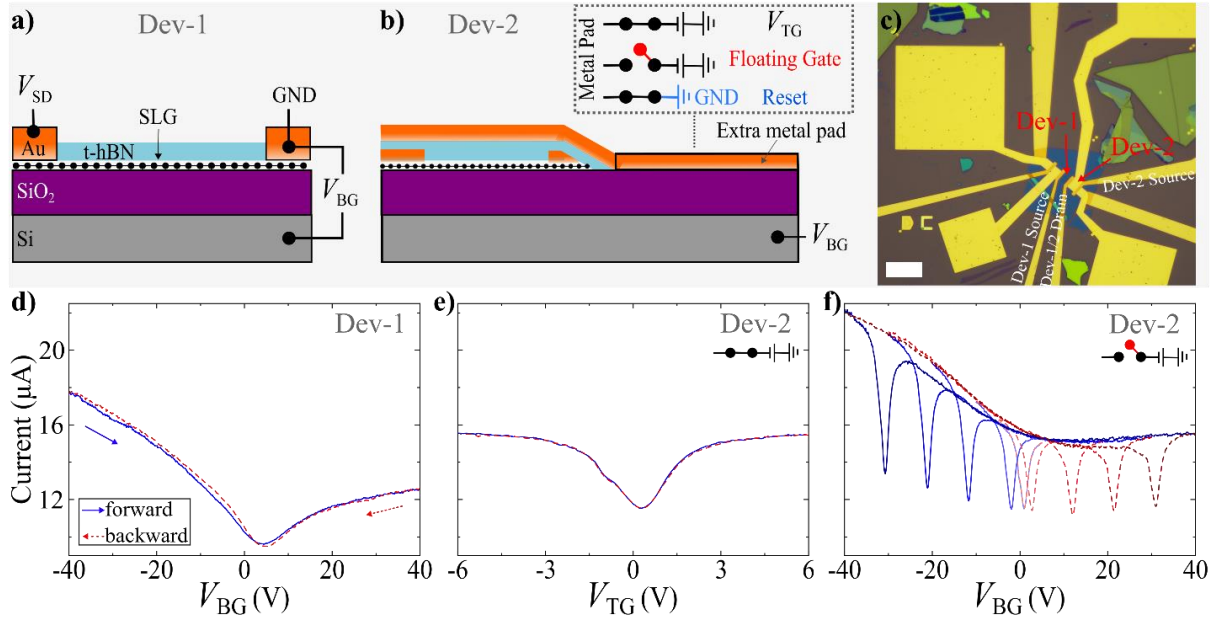

**Figure S4. a)** Experimental scheme of a capped hBN/SLG GFET operated solely in a back-gate configuration, Dev-1. **b)** Cross-section view of the experimental scheme for a double-gated GFET, Dev-2, highlighting the three working principles: top-gate bias ($V_{TG}$); floating gate (FG); or reset. **c)** Optical image of a representative capped hBN/SLG GFET, indicating both devices, Dev-1 and Dev-2. The scale bar represents 20 μm. **d)** Transfer characteristics of the Dev-1 under $V_{BG}$ loop. **e-f)** Transfer curves of the Dev-2 under $V_{TG}$ and $V_{BG}$ loops. Test conditions are fixed $V_{DS}$ = 100 mV, same bias sweep rate, and in vacuum at 300 K, with an extra metal pad size of 80 × 80 μm$^2$.

Next, we focus on the hBN/SLG heterostructure and characterize the device performance as a function of the thickness of the top-hBN (t-hBN) flakes. **Figure S5a-b** shows the position of the CNP for devices with different t-hBN thicknesses in the forward (blue symbols) and backward (red symbols) sweeps under top- and back-gate bias, respectively. Like the fully encapsulated devices, **Figure S5b** shows an enhancement of the memory window for thinner t-hBN flakes. **Figure S5c** shows the threshold bias required to activate the memory window obtained from the linear fit.

28differ by the presence of the FG electrode: Dev-1 is only back-gated, while Dev-2 is dual-gated. The transfer curves obtained from the individual devices are shown in **Figure S4d-f** and display a rather different electrical behavior. While the device without FG shows no electrical hysteresis, the one with FG confirms the existence of a very large split in the CNP position for the $V_{BG}$ loop. Finally, Dev-2 shows no electrical hysteresis for $V_{TG}$ loops, as expected.

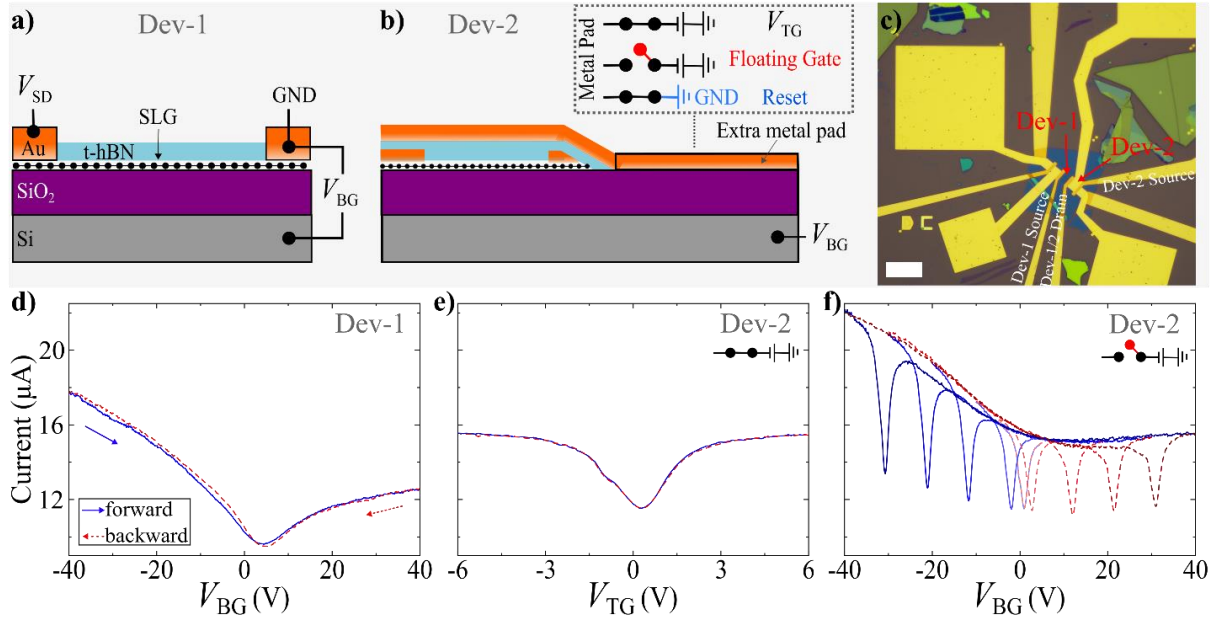

**Figure S4. a)** Experimental scheme of a capped hBN/SLG GFET operated solely in a back-gate configuration, Dev-1. **b)** Cross-section view of the experimental scheme for a double-gated GFET, Dev-2, highlighting the three working principles: top-gate bias ($V_{TG}$); floating gate (FG); or reset. **c)** Optical image of a representative capped hBN/SLG GFET, indicating both devices, Dev-1 and Dev-2. The scale bar represents 20 μm. **d)** Transfer characteristics of the Dev-1 under $V_{BG}$ loop. **e-f)** Transfer curves of the Dev-2 under $V_{TG}$ and $V_{BG}$ loops. Test conditions are fixed $V_{DS}$ = 100 mV, same bias sweep rate, and in vacuum at 300 K, with an extra metal pad size of 80 × 80 μm$^2$.

Next, we focus on the hBN/SLG heterostructure and characterize the device performance as a function of the thickness of the top-hBN (t-hBN) flakes. **Figure S5a-b** shows the position of the CNP for devices with different t-hBN thicknesses in the forward (blue symbols) and backward (red symbols) sweeps under top- and back-gate bias, respectively. Like the fully encapsulated devices, **Figure S5b** shows an enhancement of the memory window for thinner t-hBN flakes. **Figure S5c** shows the threshold bias required to activate the memory window obtained from the linear fit.

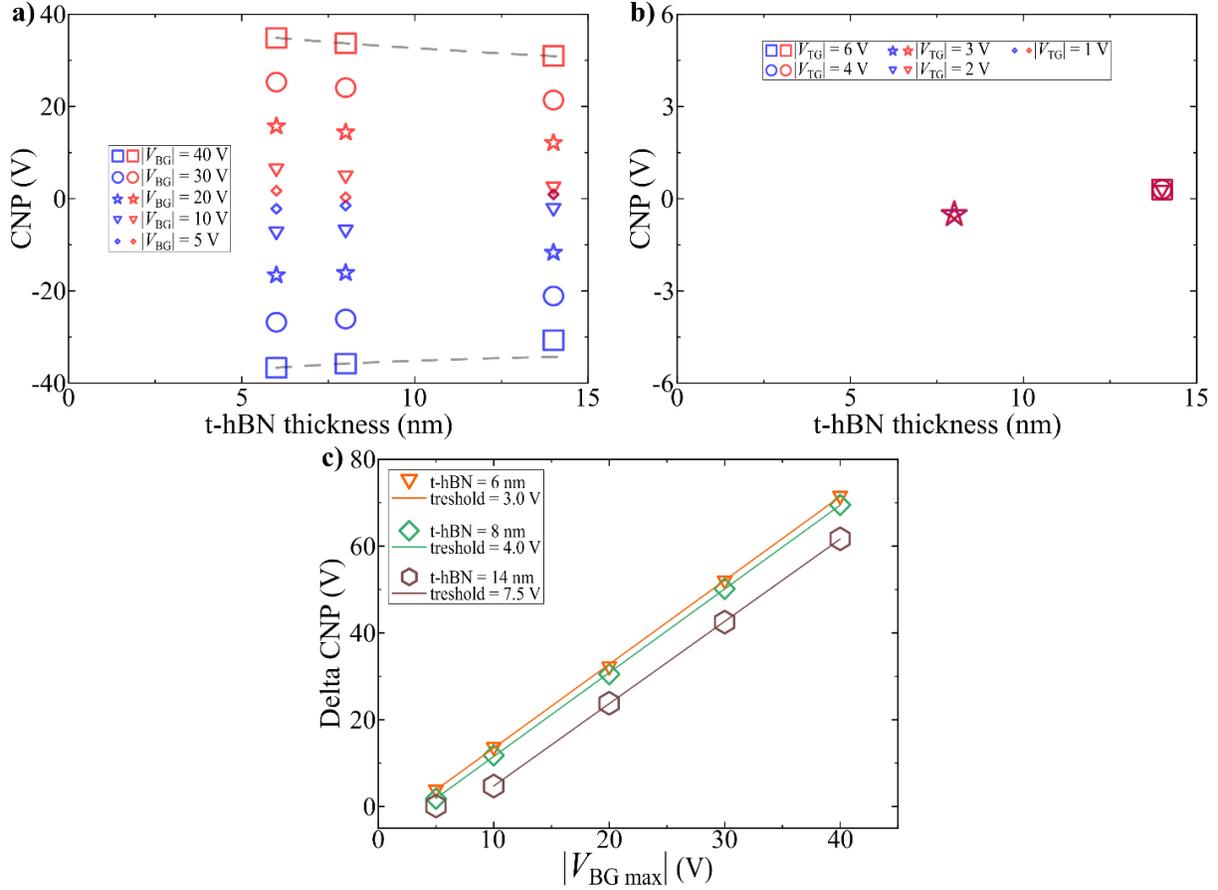

**Figure S5. a-b)** Position of CNP for capped GFETs in the forward (blue symbols) and backward (red symbols) $V_{TG}$ and $V_{BG}$ sweeps as a function of top-hBN (t-hBN) thicknesses, respectively. **c)** Memory window extracted from (a) for different devices and linear fits.

### S.5 – EBL Acceleration Independence

The electrical hysteresis discussed in our work could eventually be associated with charge trapping at the $SiO_2$ layer, induced by electron beam (e-beam) irradiation during the lithography (EBL) of the top electrode, as described in Reference[2]. To investigate this, we fabricated additional devices using different e-beam acceleration voltages during the EBL step: 5 kV and 30 kV, and the results are shown in **Figure S6**. The dashed gray line indicates the average data for 20 kV. The data indicate no correlation between the memory window and e-beam acceleration, with a similar memory window for all cases, thus suggesting that possible charges[2] at the $SiO_2$ layer are not the main effect in our devices.

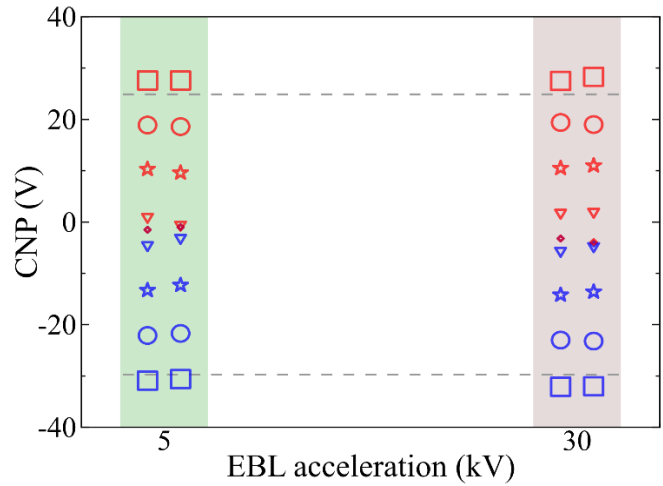

**Figure S6.** Position of the CNP for GFETs in the forward (blue symbols) and backward (red symbols) $V_{BG}$ sweeps fabricated with different EBL acceleration voltage, but similar dose (300 µC/cm$^2$). Test conditions are fixed $V_{DS}$ =100 mV, extra metal pad size is 200 × 200 µm$^2$, and in vacuum at 300 K for all GFETs. t-hBN (b-hBN) thickness are: 34 nm, 46 nm, (45 nm, 56 nm), and 34 nm, 36 nm (43 nm, 48 nm), for devices fabricated under 5 kV and 30 kV acceleration, respectively.

### S.6 – Extra Metal Pads Influence

**Figure S7** brings optical images of different GFETs fabricated and measured with distinct FG areas. **Figure S1** shows extra optical images of the devices measured.

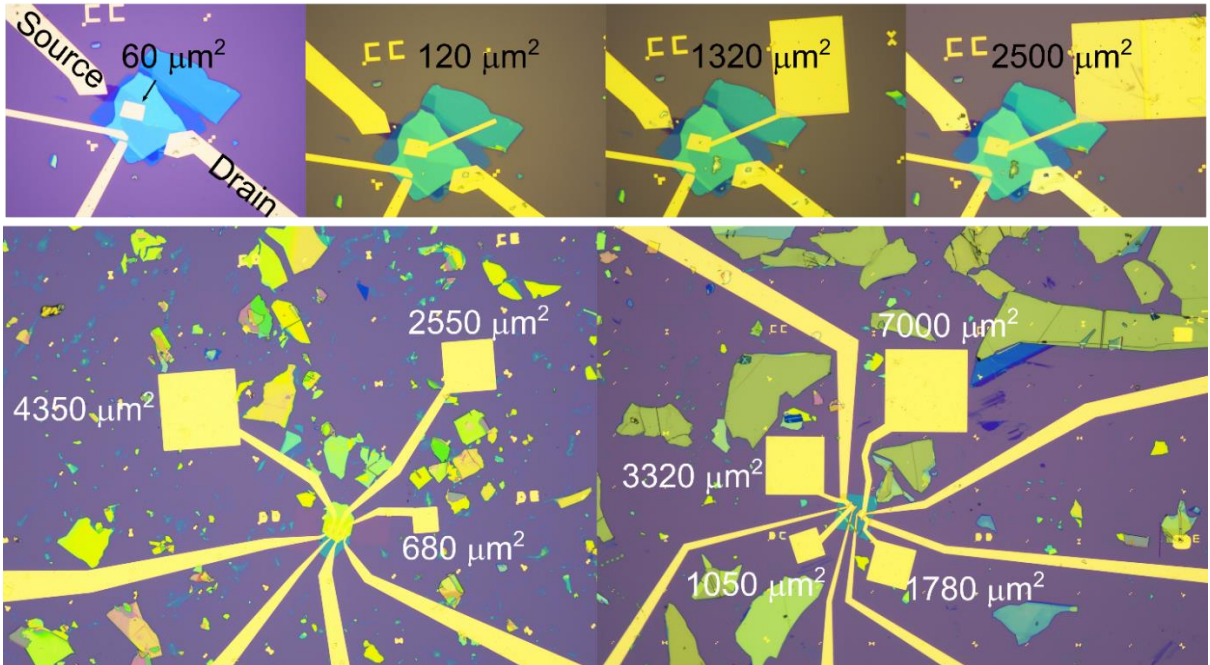

**Figure S7.** Optical images of GFETs fabricated with FG electrodes of different areas. The FG

electrodes (areas) are indicated in all images, and all other electrodes are source and drain for each device.

**Figure S8** plots the transfer curves of a representative GFET fabricated and measured with distinct FG areas. Here, the FG areas were defined over the same device by performing sequential fabrication steps, intercalated by electrical characterization, as shown in the top panels of **Figure S7**.

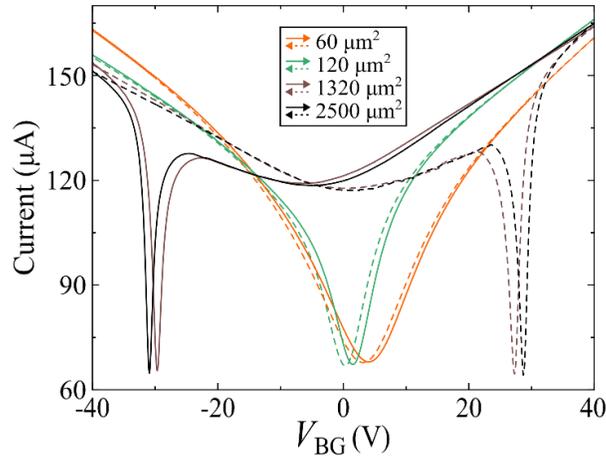

From this analysis, one can clearly observe the drastic change in the shape and position of the CNP for different FG areas.

**Figure S8.** Transfer curves of a GFET fabricated with distinct FG areas: 60, 120, 1320, 2500 µm$^2$. The electrical characterization was performed after each new FG fabrication step (see Figure S7). Test conditions are fixed $V_{DS}$ =100 mV and in vacuum at 300 K.

## S.7 – Stability of Memory Devices Over 1 Year

**Figure S9a-b** plots the transfer curves of a representative GFET fabricated and measured after fabrication and after 1 year kept in the laboratory. The data reveals negligible changes over time, supporting the robustness of the encapsulation step for reliable memory device operation.

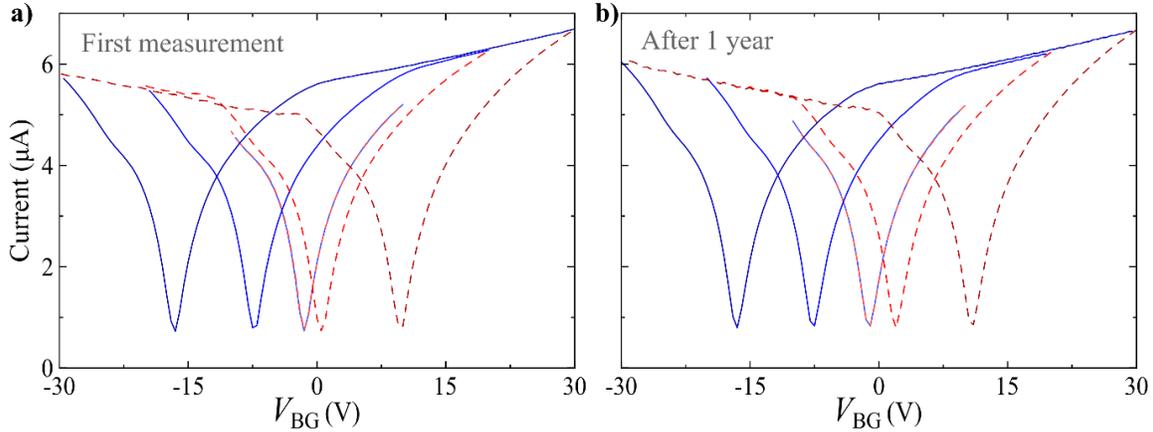

**Figure S9. a-b)** Transfer curves of a GFET measured after fabrication and after 1 year kept in laboratory conditions, respectively. The thickness of the top-hBN is 35 nm, and the test conditions are fixed $V_{DS}$ = 10 mV and in vacuum at 300 K.

## S.8 - Electrical Characteristics at Cryogenic Temperatures and Different Sweep Rates

**Figure S10a** plots the transfer curves of a representative GFET with an FG measured at different cryogenic temperatures, whereas **Figure S10b** shows the curves for different $V_{BG}$ sweep rates. The results indicate minor effects at cryogenic temperatures and for different loop periods. This is a clear difference compared to volatile graphene memory devices[3,4].

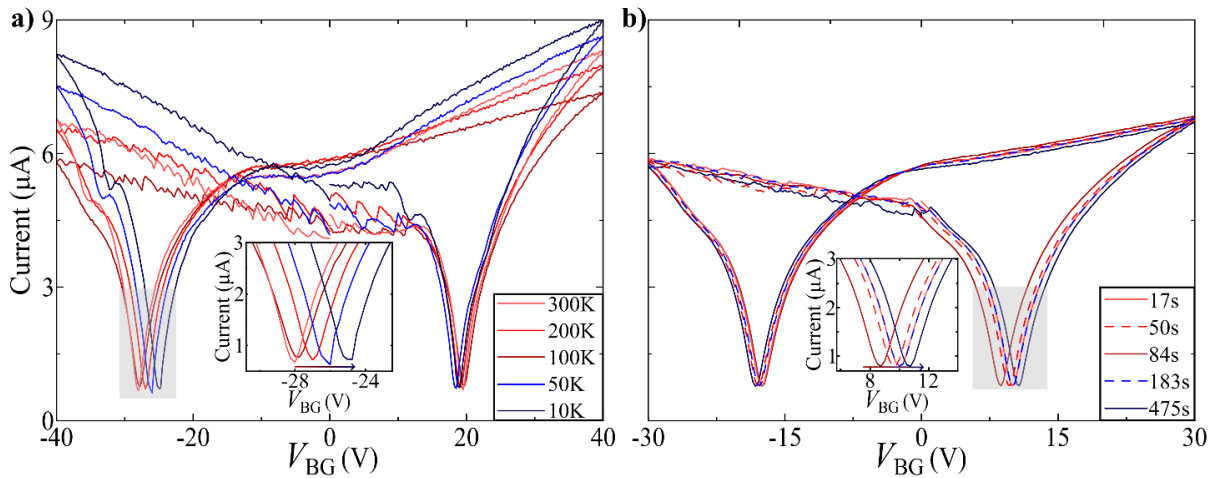

**Figure S10. a-b)** Transfer curves of a GFET measured at different temperatures and $V_{BG}$ loop periods, respectively. Test conditions are fixed $V_{DS}$ =10 mV, 200 x 200 $\mu m^2$, and in vacuum.

## S.9 – Performance Comparison

**Table S1.** Comparison of the performance of our floating-gate memory devices with other devices based on van der Waals heterostructures.

| Programming Efficiency $\left(\frac{\Delta CNP}{2 \times |V_{BG, max}|}\right)$ (%) | # Cycles for memory test | 2D Channel | Device Temperature Operation | FG Device Structure | Thickness of the tunneling layer | Reference |
|---|---|---|---|---|---|---|
| 90% | 9800 | SLG | 10 – 300 K | Au as Top FG | 6 nm | Our Work |
| 75% | 2800 | $MoS_2$ | 10 – 300 K | *MLG as bottom FG | 7.5 nm | Ref.[6] |
| 75% | x | SLG | 22 – 300 K | $MoS_2$ as bottom FG | x | Ref.[9] |
| 60% | $1 \times 10^5$ | $MoS_2$ | 300 K | Au as Top FG | 10 nm | Ref.[5] |
| 50% | 100 | SLG | 300 – 475 K | *$MoS_2$ as bottom FG | 6 nm | Ref.[7] |
| 40% | 4500 | $MoS_2$ | 300 – 400 K | SLG as bottom FG | 10 nm | Ref.[8] |

*MLG – Multi-Layer Graphene;
*$MoS_2$ – 5 nm $MoS_2$ tick;
x – values not reported;

## REFERENCES

(1) Viti, L.; Cadore, A. R.; Yang, X.; Vorobiev, A.; Muench, J. E.; Watanabe, K.; Taniguchi, T.; Stake, J.; Ferrari, A. C.; Vitiello, M. S. Thermoelectric Graphene Photodetectors with Sub-Nanosecond Response Times at Terahertz Frequencies. *Nanophotonics* **2020**, *10* (1), 89–98. https://doi.org/10.1515/nanoph-2020-0255.

(2) Shi, W.; Kahn, S.; Jiang, L.; Wang, S. Y.; Tsai, H. Z.; Wong, D.; Taniguchi, T.; Watanabe, K.; Wang, F.; Crommie, M. F.; Zettl, A. Reversible Writing of High-Mobility and High-Carrier-Density Doping Patterns in Two-Dimensional van Der Waals Heterostructures. *Nat Electron* **2020**, *3* (2), 99–105. https://doi.org/10.1038/s41928-019-0351-x.

(3) Cadore, A. R.; Mania, E.; Watanabe, K.; Taniguchi, T.; Lacerda, R. G.; Campos, L. C. Thermally Activated Hysteresis in High Quality Graphene/h-BN Devices. *Appl Phys Lett* **2016**, *108* (23), 233101. https://doi.org/10.1063/1.4953162.

(4) Wang, H.; Wu, Y.; Cong, C.; Shang, J.; Yu, T. Hysteresis of Electronic Transport in Graphene Transistors. *ACS Nano* **2010**, *4* (12), 7221–7228. https://doi.org/10.1021/nn101950n.

(5) Wang, S.; He, C.; Tang, J.; Lu, X.; Shen, C.; Yu, H.; Du, L.; Li, J.; Yang, R.; Shi, D.; Zhang, G. New Floating Gate Memory with Excellent Retention Characteristics. *Adv Electron Mater* **2019**, *5* (4). https://doi.org/10.1002/aelm.201800726.

(6) Liu, L.; Liu, C.; Jiang, L.; Li, J.; Ding, Y.; Wang, S.; Jiang, Y.-G.; Sun, Y.-B.; Wang, J.; Chen, S.; Zhang, D. W.; Zhou, P. Ultrafast Non-Volatile Flash Memory Based on van Der Waals Heterostructures. *Nat Nanotechnol* **2021**, *16* (8), 874–881. https://doi.org/10.1038/s41565-021-00921-4.

(7) Choi, M. S.; Lee, G.-H.; Yu, Y.-J.; Lee, D.-Y.; Lee, S. H.; Kim, P.; Hone, J.; Yoo, W. J. Controlled Charge Trapping by Molybdenum Disulphide and Graphene in Ultrathin Heterostructured Memory Devices. *Nat Commun* **2013**, *4*, 1624. https://doi.org/10.1038/ncomms2652.

(8) Wang, H.; Guo, H.; Guzman, R.; JiaziLa, N.; Wu, K.; Wang, A.; Liu, X.; Liu, L.; Wu, L.; Chen, J.; Huan, Q.; Zhou, W.; Yang, H.; Pantelides, S. T.; Bao, L.; Gao, H. J. Ultrafast Non-Volatile Floating-Gate Memory Based on All-2D Materials. *Advanced Materials* **2024**, *36* (24). https://doi.org/10.1002/adma.202311652.